\documentclass[useAMS,usenatbib]{mn2e}

\usepackage{amssymb,amsmath}
\usepackage{graphicx}
\usepackage{float}
\usepackage{multirow}
\usepackage{aas_macros}
\usepackage{fixltx2e}
\usepackage[usenames]{color}

\newcommand\msun {M$_{\odot}$}

\def\lum{erg s$^{-1}$}
\def\xmm{XMM-{\it Newton}}
\def\chan{{\it Chandra}}

\title[NIR counterparts of ULXs]{Near-infrared counterparts of ultraluminous X-ray sources}
\author[M. Heida et al.]
{M. Heida$^{1,2}$, P. G. Jonker$^{1,2,3}$, M. A. P. Torres$^{1}$, E. Kool$^1$, M. Servillat$^{4,3}$,
\newauthor T. P. Roberts$^5$, P. J. Groot$^2$, D. J. Walton$^6$, D. Moon$^7$, F. A. Harrison$^6$\\
$^1$SRON Netherlands Institute for Space Research, Sorbonnelaan 2, 3584 CA Utrecht, the Netherlands\\
$^2$Department of Astrophysics/IMAPP, Radboud University Nijmegen, P.O. Box 9010, 6500 GL Nijmegen, The Netherlands\\
$^3$Harvard-Smithsonian Center for Astrophysics, 60 Garden Street, Cambridge, MA 02138, USA\\
$^4$Laboratoire AIM (CEA/Irfu/SAp, CNRS/INSU, Universit\'e Paris Diderot), CEA Saclay, 91191 Gif-sur-Yvette, France\\
$^5$Department of Physics, University of Durham, South Road, Durham DH1 3LE, United Kingdom\\
$^6$Space Radiation Laboratory, California Institute of Technology, Pasadena, CA 91125, USA\\
$^7$Department of Astronomy and Astrophysics, University of Toronto, Toronto, ON M5S 3H4, Canada\\
}

\begin{document}

\maketitle

\begin{abstract}
In this paper we present the results of the first systematic search for counterparts to nearby ultraluminous X-ray sources (ULXs) in the near-infrared (NIR). We observed 62 ULXs in 37 galaxies within 10 Mpc and discovered 17 candidate NIR counterparts. 
The detection of 17 out of 62 ULX candidates points to intrinsic differences between systems that show and those that do not show infrared emission. For six counterparts we conclude from the absolute magnitudes and -- in some cases -- additional information such as morphology and previously reported photometric or spectroscopic observations, that they are likely background active galactic nuclei or ULXs residing in star clusters. Eleven counterparts have absolute magnitudes consistent with them being single red supergiant stars. Alternatively, these systems may have larger accretion discs that emit more NIR light than the systems that we do not detect. Other scenarios such as emission from a surrounding nebula or from a compact radio jet are also possible, although for Holmberg II X-1 the NIR luminosity far exceeds the expected jet contribution. 
The eleven possible red supergiant counterparts are excellent candidates for spectroscopic follow-up observations. This may enable us to measure the mass function in these systems if they are indeed red supergiant donor stars where we can observe absorption lines.
\end{abstract}


\section{Introduction}
Ultraluminous X-ray sources (ULXs) are point-like, off-nuclear X-ray sources with an X-ray luminosity above $10^{39}$ \lum \citep{feng11}, the Eddington luminosity of a $\sim$10 \msun\ black hole. The most luminous example reaches $10^{42}$ \lum~(HLX-1, \citealt{farrell09}). Several scenarios have been proposed to explain the high luminosities of these sources. Geometrical beaming could boost the observed luminosity by up to a factor $\sim$10 (\citealt{king01}). Genuine super-Eddington accretion is another option \citep{begelman02,moon03}. Alternatively, if these systems contain black holes that are more massive than 100 \msun (which is our working definition of intermediate mass black holes, IMBHs), that would naturally explain the high luminosities. 

Observations indicate that ULXs at lower and higher luminosity may belong to different populations. The luminosity functions of ULXs and fainter extra-galactic X-ray binaries show a break around $2 \times 10^{40}$ \lum{} (\citealt{swartz11,mineo12}). At luminosities below this break, the identification of an `ultraluminous' X-ray spectral state (e.g., \citealt{gladstone09}) added observational support to the idea that the majority of these ULXs are stellar-mass black holes accreting at super-Eddington rates, which are rarely observed in Galactic black hole X-ray binaries for reasons currently unknown. Alternatively, for a low-metallicity star it is possible to leave a black hole of $\lesssim 70$ M$_\odot$ \citep{belczynski10}. Accretion onto such a black hole could explain ULXs with luminosities up to a few times $10^{40}$ erg/s \citep{zampieri09}. In this scenario only the brightest ULXs, with X-ray luminosities $\gtrsim 2 \times 10^{40}$ \lum, remain as plausible IMBH candidates. Such a seemingly convincing case for an IMBH is provided by the ULX ESO~243--49 X--1 \citep{farrell09}. 

Ultimately, the most reliable way to determine the nature of ULXs is to obtain dynamical mass measurements of their black hole masses using radial velocity curves of their companion stars, as well as measurements of the system inclination and binary mass ratio. Dynamical mass measurements of Galactic binary systems usually rely on optical spectroscopic observations of the motion of the companion star. Unfortunately, most attempts to use the same technique for ULXs have not met with success (c.f. \citealt{roberts11,liu12,cseh13a}, however, see the recent work of \citealt{liu13}). The reason for this failure is that most optical counterparts to ULXs are faint, and, in the cases where it has been possible to perform spectroscopic studies, the spectra show strong emission lines from the accretion disc, the irradiated companion star and/or from the surrounding nebulae but no absorption features from a stellar component. The emission lines can not be used for reliable dynamical mass measurements because their radial velocities vary erratically and on timescales that may be unrelated to the orbital period of the system. 

Searches for photospheric lines from the donor stars in ULXs have thus far concentrated on the blue part of the spectrum because ULX donor stars were expected to be blue supergiants. This expectation is based on the fact that many ULXs are located in or near young star clusters (e.g. \citealt{fabbiano01,roberts02,gao03,swartz09,voss11}) and on the blue colours of many ULX counterparts (e.g. \citealt{roberts01,liu02}). However, these blue colours are also consistent with emission from an irradiated accretion disc \citep{soria12,grise12}. And if ULX donor stars are indeed massive young stars, some of them may very well be Wolf-Rayet stars (like the counterpart to M 101 ULX-1, \citealt{liu13}) or red supergiants (\citealt{copperwheat05,copperwheat07,patruno08}).  

Red supergiant (RSG) stars are intrinsically bright in the NIR (M${_V} \sim -6$, V--H$\sim 4$, H--K$\sim 0$) in contrast with the blue supergiants (M${_V} \sim -6$, V--H$\sim 0$, H--K$\sim 0$; \citealt{elias85,drilling00}), and can outshine the accretion disc in that part of the spectrum. A Galactic analog is GRS1915-105, a black hole X-ray binary with a red giant companion star (not an RSG) with an orbital period of 33.5 days. A dynamical mass measurement of the black hole in this system was obtained from radial velocity studies in the NIR while the source was bright in X-rays, showing that irradiation of the companion star has little effect on long orbital period systems like this \citep{steeghs13}. A Roche-lobe overflowing red supergiant would have an even larger orbit than the red giant in GRS1915-105, so irradiation is not likely to produce strong effects in the companion atmosphere  (\citealt{copperwheat05,copperwheat07}). With the largest telescopes currently available, it is possible to perform time-resolved spectroscopy for sources with $H$- and $K$-band magnitudes up to $\sim$ 20. This corresponds to the typical apparent magnitude of an RSG at a distance of $\sim$ 10 megaparsec (Mpc). 

In this paper we present the results of our search for candidate RSG companions to ULXs within 10 Mpc. The sample consists of sources from several catalogs of ULX candidates. 
We describe the sample in Section 2, the observations in Section 3 and the data reduction Sections 4. The results, including a table summarizing all detected counterparts and limiting magnitudes for the non-detections, can be found in Section 5. In Section 6 we discuss our findings.

\section{Sample}
Our sample consists mainly of ULX candidates from the catalogs of \citet{swartz04, swartz11} and \citet{walton11b}, supplemented with additional targets from other sources (see Tables \ref{targettabel1} and \ref{targettabel2}). We only targeted ULX candidates within a distance of 10 Mpc, since this is the maximum distance at which it is still viable to take a NIR spectrum of an RSG with existing large telescopes. ULX candidates in very crowded regions, such as edge-on spiral galaxies and fields close to galactic centers, were excluded from the sample. 

Five candidates had no reported accurate X-ray positions when we started our observing campaign. For these sources we determined an accurate X-ray position based on archival \chan{}/ACIS observations. We used the {\sc ciao} \citep{fruscione06} task ACIS\_PROCESS\_EVENTS to reprocess the event files if the observations were conducted in very faint mode. Then we determined accurate positions for the ULX candidates using WAVDETECT. The sources and \chan{} observation ID's are listed in Table \ref{chantabel}. For the other sources X-ray positions and uncertainties were taken from the literature (see Table \ref{targettabel1}). Quoted error radii on the source location are at the 95\% confidence level (2-$\sigma$).

The complete sample consists of 62 ULX candidates in 37 galaxies. We have collected recent distance measurements for these galaxies where available to compute absolute magnitudes and absolute magnitude limits for the NIR counterparts, assuming they are at the same distance as their host galaxies (see Table \ref{logtabel}). These are generally different from the distance measurements used to compile catalogs of ULX candidates. For example, \citet{walton11b} use only distances from \citet{1988ngc..book.....T}. Comparing these to the newer distance measurements showed that if there are significant differences, these older measurements generally underestimate the distance. Hence the true X-ray luminosities of these sources only go up with the new distance measurements and all sources in our sample are still ULX candidates.

 \begin{table*}
  \caption{The ULX candidates of which we obtained NIR images, with their designation from Simbad, their best determined X-ray position with the radius of the 95\% confidence error circle, and the catalog (`this work' for sources for which we determined the position) and satellite that provided the position.}\label{targettabel1}
  \begin{tabular}{llccccc}
  \hline
Galaxy & ULX name (SIMBAD) & R.A. (h:m:s) & Dec. (d:m:s) & Error & Source & Satellite \\
\hline
NGC 253 & RX J004722.4-252051 & 00:47:22.59 & -25:20:50.9 & 1.0'' & \citealt{liu11} & \chan \\
NGC 253 & RX J004742.5-251501 & 00:47:42.76 & -25:15:02.2 & 1.0'' & \citealt{liu11} & \chan \\
M74 & [KKG2005] M74 X-1 & 01:36:51.06 & 15:45:46.8 & 0.7'' & \citealt{liu11} & \chan \\
M74 & XMMU J013636.5+155036 & 01:36:36.5 & 15:50:36.3 & 0.8'' & \citealt{lin12}  & \xmm \\
NGC 855 & [SST2011] J021404.08+275239.5 & 02:14:04.09 & 27:52:39.4 & 1.0'' & \citealt{liu11} & \chan \\
NGC 925 & [SST2011] J022721.52+333500.7 & 02:27:21.52 & 33:35:00.8 & 1.0'' & \citealt{liu11} & \chan \\
NGC 925 & [SST2011] J022727.53+333443.0 & 02:27:27.53 & 33:34:42.9 & 1.0'' & \citealt{liu11} & \chan \\
NGC 1058 & XMMU J024323.5+372038 & 02:43:23.27 & 37:20:42.1 & 0.7'' & this work & \chan \\
NGC 1313 & RX J0318.3-6629 & 03:18:20.00 & -66:29:10.9 & 1.0'' & \citealt{liu11} & \chan \\
IC342 & XMMU J034555.6+680455 & 03:45:55.61 & 68:04:55.3 & 0.25'' & \citealt{feng08} & \chan \\
IC342 & [SST2011] J034615.64+681112.2 & 03:46:15.61 & 68:11:12.8 & 0.4'' & \citealt{lin12} & \xmm \\
NGC 1637 & [IWL2003] 68 & 04:41:32.97 & -02:51:26.8 & 1.0'' & \citealt{liu11} & \chan \\
NGC 2403 & CXOU J073625.6+653539 & 07:36:25.55 & 65:35:40.0 & 0.7'' & \citealt{swartz04} & \chan \\
NGC 2403 & 2E 0732.2+6546 & 07:37:02.33 & 65:39:35.0 & 0.7'' & this work & \chan \\
NGC 2500 & [SST2011] J080148.10+504354.6 & 08:01:48.11 & 50:43:54.6 & 1.0'' & \citealt{liu11} & \chan \\
NGC 2500 & CXO J080157.8+504339 & 08:01:57.84 & 50:43:39.4 & 1.0'' & \citealt{liu11} & \chan \\
Holmberg II & Holmberg II X-1 & 08:19:28.99 & 70:42:19.4 & 0.7'' & \citealt{swartz04} & \chan \\
NGC 2903 & CXOU J093206.2+213058 & 09:32:06.19 & 21:30:58.9 & 0.6'' & Watson et al. 2014, in prep & \xmm \\
Holmberg I & [WMR2006] Ho I XMM1 & 09:41:30.15 & 71:12:35.7 & 1.0'' & Watson et al. 2014, in prep & \xmm \\
Holmberg I & 1WGA J0940.0+7106 & 09:39:59.44 & 71:06:40.2 & 1.0'' & Watson et al. 2014, in prep & \xmm \\
Holmberg I &  [WMR2006] Ho I XMM3 & 09:42:06.36 & 71:04:40.0 & 1.4'' & Watson et al. 2014, in prep & \xmm \\
M81 & [LM2005] NGC 3031 ULX1 & 09:55:32.95 & 69:00:33.6 & 1.0'' & \citealt{liu11} & \chan \\
Holmberg IX & Holmberg IX X-1 & 09:57:53.31 & 69:03:48.1 & 1.0'' & \citealt{liu11} & \chan \\
NGC 3184 & CXOU J101812.0+412421 & 10:18:12.05 & 41:24:20.7 & 0.7'' & \citealt{swartz04} & \chan \\
NGC 3239 & [SST2011] J102506.98+170947.2 & 10:25:06.98 & 17:09:47.2 & 1.0'' & \citealt{liu11} & \chan \\
NGC 3239 & [SST2011] J102508.20+170948.3 & 10:25:08.20 & 17:09:48.4 & 1.0'' & \citealt{liu11} & \chan \\
NGC 3486 & XMMU J110022.4+285818 & 11:00:22.27 & 28:58:16.9 & 0.7'' & this work & \chan \\
NGC 3521 & [SST2011] J110545.62+000016.2 & 11:05:45.63 & 00:00:16.5 & 1.0'' & \citealt{liu11} & \chan \\
NGC 3621 & [GSE2009] B & 11:18:15.16 & -32:48:40.6 & 0.7$^{\dag}$ & \citealt{gliozzi09} & \chan \\
NGC 3623 & [LB2005] NGC 3623 ULX1 & 11:18:58.54 & 13:05:30.9 & 1.0'' & \citealt{liu11} & \chan \\
NGC 3627 & [SST2011] J112020.90+125846.6 & 11:20:20.89 & 12:58:46.0 & 1.0'' & \citealt{liu11} & \chan \\
NGC 3627 & [SST2011] J112018.32+125900.8 & 11:20:18.31 & 12:59:00.3 & 1.0'' & \citealt{liu11} & \chan \\
NGC 3628 & CXOU J112037.3+133429 & 11:20:37.37 & 13:34:29.2 & 1.0'' & \citealt{liu11} & \chan \\
NGC 4136 & CXOU J120922.6+295551 & 12:09:22.58 & 29:55:50.6 & 1.0'' & \citealt{liu11} & \chan \\
NGC 4136 & [SST2011] J120922.18+295559.7 & 12:09:22.19 & 29:55:59.7 & 1.0'' & \citealt{liu11} & \chan \\
NGC 4204 & [SST2011] J121510.91+203912.4 & 12:15:10.91 & 20:39:12.4 & 1.0'' & \citealt{liu11} & \chan \\
NGC 4258 & RX J121844.0+471730 & 12:18:43.88 & 47:17:31.7 & 1.0'' & \citealt{liu11} & \chan \\
NGC 4258 & RX J121857.7+471558 & 12:18:57.85 & 47:16:07.4 & 1.0'' & \citealt{liu11} & \chan \\
NGC 4258 & [WMR2006] NGC4258 XMM1 & 12:18:47.66 & 47:20:54.7 & 0.8'' & Watson et al. 2014, in prep & \xmm \\
NGC 4258 & RX J121845.6+472420 & 12:18:45.51 & 47:24:20.2 & 1.0'' & Watson et al. 2014, in prep & \xmm \\
NGC 4395 & IXO 53 & 12:26:01.53 & 33:31:30.6 & 1.0'' & \citealt{liu11} & \chan \\
NGC 4449 & RX J122818.0+440634 & 12:28:17.83 & 44:06:33.9 & 1.1'' & \citealt{liu11} & \chan \\
NGC 4559 & RX J123551+27561 & 12:35:51.71 & 27:56:04.1 & 0.7'' & \citealt{swartz04} & \chan \\
NGC 4618 & [SST2011] J124129.14+410757.7 & 12:41:29.14 & 41:07:57.7 & 0.7''$^{\dag}$ & \citealt{swartz11} & \chan \\
NGC 5128 & 1RXH J132519.8-430312 & 13:25:19.87 & -43:03:17.1 & 1.0'' & \citealt{kraft01} & \chan \\
NGC 5128 & CXOU J132518.2-430304 & 13:25:18.24 & -43:03:04.5 & 0.4'' & \citealt{sivakoff08} & \chan \\
NGC 5204 & CXOU J132938.6+582506 & 13:29:38.62 & 58:25:05.6 & 1.0'' & \citealt{liu11} & \chan \\
M51 & RX J132943+47115 & 13:29:43.31 & 47:11:34.8 & 0.7'' & \citealt{terashima06} & \chan \\
M51 & XMMU J132950.7+471153 & 13:29:50.68 & 47:11:55.2 & 0.7''  & \citealt{terashima06} & \chan \\
M51 & XMMU J132953.3+471040 & 13:29:53.31 & 47:10:42.5 & 0.7'' & \citealt{terashima06}  & \chan \\
M51 & RX J132954+47145 & 13:29:53.72 & 47:14:35.7 & 0.7'' & \citealt{terashima06}  & \chan \\
M51 & XMMU J132957.6+471047 & 13:29:57.57 & 47:10:48.3 & 0.7'' & \citealt{terashima04} & \chan \\
M51 & RX J133001+47137 & 13:30:01.01 & 47:13:43.9 & 0.7'' & \citealt{terashima06} & \chan \\
M51 & RX J133006+47156 & 13:30:06.00 & 47:15:42.3 & 0.7'' & \citealt{terashima06} & \chan \\
M51 & RX J133007+47110 & 13:30:07.55 & 47:11:06.1 & 0.7'' & \citealt{terashima06} & \chan \\
NGC 5408 & NGC 5408 X-1 & 14:03:19.63 & -41:22:58.7 & 1.2'' & \citealt{kaaret03} & \chan \\
NGC 5474 & NGC 5474-X1 & 14:04:59.74 & 53:38:09.0 & 1.0'' & \citealt{liu11} & \chan \\
M101 & CXOU J140332.3+542103 & 14:03:32.38 & 54:21:03.0 & 0.7'' & \citealt{swartz04} & \chan \\
M101 & 2E 1402.4+5440 & 14:04:14.28 & 54:26:03.6 & 0.7'' & \citealt{swartz04} & \chan \\
M101 & 2XMM J140248.0+541350 & 14:02:48.19 & 54:13:50.7 & 0.7'' & this work & \chan \\
M101 & CXOU J140314.3+541807 & 14:03:14.33 & 54:18:06.7 & 0.7'' & this work & \chan \\
NGC 5585 & [SST2011] J141939.39+564137.8 & 14:19:39.39 & 56:41:37.8 & 1.0'' & \citealt{liu11} & \chan \\
\hline
\end{tabular}
Notes - $^{\dag}$: No positional error was given in the cited article, we assume the standard \chan{} bore-sight error of 0.6'' (90\% confidence) and a localization error of less than 0.1'' (90\% confidence), which combine into an 0.7'' total error (95\% confidence).
\end{table*}

\begin{table*}
  \caption{Entries (marked with $\bullet$) and/or names of our sources in other catalogs. Sources are listed in the same order as in Table \ref{targettabel1}. (1): \citet{swartz04}, (2): \citet{liu05}, (3): \citet{liu05b}, (4): \citet{winter06}, (5): \citet{walton11b},(6): \citet{swartz11}, (7): \citet{barnard10}.}\label{targettabel2}
  \begin{tabular}{llccccccc}
  \hline
ULX name (short) & (1) & (2) & (3) & (4) & (5) & (6) & Other \\
\hline
J004722-252051 &  & NGC 253 X9 &  & NGC 253 XMM2 & $\bullet$ &  & NGC 253 ULX1$^7$ \\
J004742-251501 &  & NGC 253 X6 &  & NGC 253 XMM6 &  &  & NGC 253 ULX3$^7$ \\
M74 X-1 & $\bullet$ &  & NGC 628 ULX1 &  & $\bullet$ & & CXOU J013651.1+154547 \\
J013636+155036 &  &  & NGC 628 ULX2 &  & $\bullet$ &  &  \\
J021404+275239  &  &  &  &  &  & $\bullet$ & CXOU J021404.0+275239 \\
J022721+333500  &  &  &  &  &  & $\bullet$ & CXOU J022721.5+333500 \\
J022727+333443 &  &  &  &  &  & $\bullet$ & CXOU J022727.5+333442 \\
J024323+372038 &  &  & NGC 1058 ULX1 &  & $\bullet$ &  & CXOU J024323.2+372042 \\
J0318-6629 &  & NGC 1313 X2 & NGC 1313 ULX1 & NGC 1313 XMM1 & $\bullet$ &  & CXOU J031820.0-662911 \\
J034555+680455 &  & PGC13826 X6 & IC 342 ULX1 & IC 342 XMM1 & $\bullet$ & $\bullet$ & IC 342 X-1 \\
J034615+681112 &  & PGC13826 X7 &  &  & $\bullet$ & $\bullet$ & 2XMM J034615.6+681112 \\
$[$IWL2003$]$ 68 &  &  &  &  &  &  & CXOU J044132.9-025126 \\
J073625+653539 & $\bullet$ & NGC 2403 X2 & NGC 2403 ULX1 & NGC 2403 XMM1 & $\bullet$ & $\bullet$ &  \\
J0732+6546 &  & NGC 2403 X3 &  & NGC 2403 XMM4 & $\bullet$ &  & CXOU J073702.3+653935 \\
J080148+504354 &  &  &  &  &  & $\bullet$ & CXOU J080148.1+504354 \\
J080157+504339 &  &  &  &  &  & $\bullet$ &  \\
Holmberg II X-1 & $\bullet$ & PGC 23324 ULX1 & Holmberg II ULX1 & Ho II XMM1 & $\bullet$ & $\bullet$ &  \\
J093206+213058 &  &  &  &  &  & $\bullet$ & 3XMM J093206.1+213058 \\
Ho I XMM1 &  &  &  & Ho I XMM1 &  &  & 3XMM J094130.1+711235 \\
J0940+7106 &  &  &  & Ho I XMM2 &  &  & 3XMM J093959.4+710640 \\
Ho I XMM3 &  &  &  & Ho I XMM3 &  &  & 3XMM J094206.3+710444 \\
NGC 3031 ULX1 & $\bullet$ &  & NGC 3031 ULX1 & M81 XMM1 &  & $\bullet$ & CXOU J095532.9+690033	 \\
Holmberg IX X-1 &  & PGC 28757 X2 & NGC 3031 ULX2 & Ho IX XMM1 &  &  & CXOU J095753.3+690348 \\
J101812+412421 & $\bullet$ &  &  &  &  & $\bullet$ &  \\
J102506+170947 &  &  &  &  &  & $\bullet$ & CXOU J102506.9+170947 \\
J102508+170948 &  &  &  &  &  & $\bullet$ & CXOU J102508.2+170948 \\
J110022+285818 &  &  &  &  & $\bullet$ &  & CXOU J110021.7+285818 \\
J110545+000016 &  &  &  &  & $\bullet$ &  & CXOU J110545.6+000016 \\
NGC3621 B &  &  &  &  &  &  &  \\
NGC 3623 ULX1 &  & NGC 3623 X2 &  &  & $\bullet$ & $\bullet$ & CXOU J111858.5+130530 \\
J112020+125846 &  & NGC 3627 X2 &  &  & $\bullet$ & $\bullet$ & CXOU J112020.8+125846 \\
J112018+125900 &  &  &  &  &  & $\bullet$ & CXOU J112018.3+125900 \\
J112037+133429 &  & NGC 3628 X2 & NGC 3628 ULX2 &  &  & $\bullet$ &  \\
J120922+295551 &  & NGC 4136 X1 & NGC 4136 ULX1 &  &  &  &  \\
J120922+295559 &  &  & NGC 4136 ULX2 &  &  & $\bullet$ & CXOU J120922.1+29555 \\
J121510+203912 &  &  &  &  &  & $\bullet$ & CXOU J121510.9+203912 \\
J121844+471730 & $\bullet$ &  &  &  & $\bullet$ & $\bullet$ & CXOU J121843.8+471731 \\
J121857+471558 & $\bullet$ &  & NGC 4258 X3 & NGC4258 XMM2 & $\bullet$ & $\bullet$ & CXOU J121857.8+471607 \\
NGC4258 XMM1 &  &  &  & NGC4258 XMM1 & $\bullet$ &  & 3XMM J121847.6+472054 \\
J121845+472420 &  &  &  &  & $\bullet$ &  & 3XMM J121845.5+472420 \\
IXO 53 &  & NGC 4395 X1 & NGC 4395 ULX1 & NGC4395 XMM1 &  & $\bullet$ & CXOU J122601.5+333130 \\
J122818+440634 & $\bullet$ & NGC 4449 X5 &  & NGC4449 XMM1 &  & $\bullet$ & CXOU J122817.8+440633 \\
J123551+27561 & $\bullet$ & NGC 4559 X5 & NGC 4559 ULX1 &  & $\bullet$ & $\bullet$ & CXOU J123551.7+275604 \\
J124129+410757 &  &  &  &  &  & $\bullet$ &  \\
J132519-430312 &  & NGC 5128 X4 & NGC 5128 ULX1 &  &  &  & CXOU J132519.9-430317 \\
J132518-430304 &  &  &  &  &  &  &  \\
J132938+582506 & $\bullet$ & NGC 5204 X1 & NGC 5204 ULX1 & NGC5204 XMM1 & $\bullet$ & $\bullet$ &  \\
J132943+47115 & $\bullet$ & NGC 5194 X4 & NGC 5194-5 ULX2 &  &  & $\bullet$ &  \\
J132950+471153 & $\bullet$ &  & NGC 5194-5 ULX3 &  &  & $\bullet$ &  \\
J132953+471040 & $\bullet$ &  & NGC 5194-5 ULX4 &  & $\bullet$ & $\bullet$ &  \\
J132954+47145 & $\bullet$ & NGC 5194 X9 & NGC 5194-5 ULX5 & M51 XMM7 &  & $\bullet$ & CXOM51 J132953.7+471436 \\
J132957+471047 & $\bullet$ &  &  & M51 XMM6 &  & $\bullet$ &  \\
J133001+47137 & & NGC 5194 X6 & NGC 5194-5 ULX7 & M51 XMM3 & $\bullet$ & $\bullet$ &  \\
J133006+47156 & & NGC 5195 X2 & NGC 5194-5 ULX9 & M51 XMM4 & $\bullet$ & $\bullet$ &  \\
J133007+47110 & $\bullet$ & NGC 5194 X8 & NGC 5194-5 ULX8 & M51 XMM2 & $\bullet$ & $\bullet$ & CXOM51 J133007.6+471106 \\
NGC 5408 X-1 &  &  & NGC 5408 ULX1 & NGC5408 XMM1 & $\bullet$ &  &  \\
NGC 5474-X1 &  &  &  &  &  & $\bullet$ & CXOU J140459.7+533808 \\
J140332+542103 & $\bullet$ & NGC 5457 X5 & NGC 5457 ULX7 &  &  & $\bullet$ &  \\
J1402+5440 & $\bullet$ & NGC 5457 X23 & NGC 5457 ULX3 & M101 XMM3 & $\bullet$ & $\bullet$ & CXOU J140414.3+542604 \\
J140248+541350 &  &  &  &  & $\bullet$ &  & CXOU J140248.1+541350 \\
J140314+541807 &  &  & NGC 5457 ULX2 & M101 XMM1 & $\bullet$ &  &  \\
J141939+564137 &  &  &  &  &  & $\bullet$ & CXOU J141939.3+564137 \\
\hline
\end{tabular}
\end{table*}

\begin{table}
 \caption{The ULX candidates for which we calculated accurate X-ray positions from archival \chan{} observations.}\label{chantabel}
 \begin{tabular}{lc}
 \hline
 ULX candidate & \chan{} obs. ID \\
 \hline
 J024323+372038 & 9579 \\
 J0732+6546 & 2014 \\
 J110022+285818 & 393 \\
 J140248+541350 & 14341 \\
 J140314+541807 & 14341 \\
 \hline
 \end{tabular}

\end{table}

\section{Observations}

We obtained observations for this project using three different telescopes\footnote{The data are publicly available at http://archive.eso.org and http://casu.ast.cam.ac.uk/casuadc/ingarch/}. In the Northern hemisphere we used the William Herschel Telescope with the Long-slit Intermediate Resolution Infrared Spectrograph (WHT/LIRIS) and the MMT with the SAO Widefield InfraRed Camera (MMT/SWIRC) to obtain $H$- and/or \textit{Ks}-band images of 51 and 16 sources, respectively. WHT/LIRIS provides a pixel scale of 0.25''/pixel and a field of view of 4.27' $\times$ 4.27', MMT/SWIRC has a pixel scale of 0.15''/pixel and a field of view of 5.12' $\times$ 5.12'. 
For targets in the Southern hemisphere we were granted 8 hours of service mode observations with the Infrared Spectrometer And Array Camera on the Very Large Telescope (VLT/ISAAC, \citealt{moorwood98}) in period 89 (program 089.D-0663(A)) and 4 hours in period 90 (program 090.D-0417(A)). This yielded \textit{Ks}-band images of 14 sources in total. We observed the 10 sources in p89 in one observing block (OB) per source. For the 4 sources in p90 we used two (shorter) OBs per source; for two sources (in NGC 1637 and NGC 3621), these OBs were executed in the same night, whereas the OBs for the two ULXs in NGC 253 were executed 14 (J004722-252051) and 42 (J004742-251501) days apart, respectively. We used the SW imaging mode of ISAAC, which gives a spatial scale of 0.148''/pixel and a field of view of 152'' $\times$ 152''. 

All observations were performed using multiple repetitions of a 5-point dither pattern. We used exposure times of 20 seconds in the $H$-band and 15 seconds in the \textit{Ks}-band with WHT/LIRIS and MMT/SWIRC, and exposure times of 6 or 10 seconds with VLT/ISAAC. The ULXs were not necessarily centered on the detectors as the pointings were selected to minimize the background from the galaxy, maximize the number of reference stars in the field and avoid very bright stars. However, if the host galaxy covered more than half of the field of view of the image we alternated between the target and an off-target blank sky field to properly subtract the sky background.

A log of the observations including the total exposure time for each source, the size of the point spread function in the image as a measure of the image quality, the distance modulus to the galaxy and the uncertainty in the astrometry is shown in Table \ref{logtabel}.

\begin{table*}
 \begin{minipage}{150mm}
 \caption{Log of the observations, with the distance modulus to the assumed host galaxy of the ULX candidates. The image number can be used to link details of the observations to the data on individual ULXs in Table \ref{resultstabel}. I.Q. is the image quality. The last column gives the 1-$\sigma$ uncertainty in the astrometry of the image.}\label{logtabel}
  \begin{tabular}{llccccccc}
  \hline
Im. & Galaxy & Filter & Instrument/ & Obs. date & Exp. time & I.Q. & Distance modulus & WCS uncertainty$^{\ddag}$\\
no. & & & Telescope & & (seconds) & (arcsec) & (magnitudes) & (arcsec) \\
  \hline
1 & NGC 253 & Ks & ISAAC/VLT & 2012-11-19$^*$ & 750 & 0.7 & $27.7 \pm 0.1^1$ & 0.26 \\
2 & NGC 253 & Ks & ISAAC/VLT & 2012-11-19$^*$ & 750 & 0.4 & $27.7 \pm 0.1^1$ & 0.48 \\
3 & M74 & Ks & ISAAC/VLT & 2012-07-06 & 260 & 1.0 & $30.0 \pm 0.4^5$ & 0.5$^{\S}$ \\
4 & M74 & Ks & LIRIS/WHT & 2012-01-01 & 7125 & 0.8 & $30.0 \pm 0.4^5$ & 0.36 \\
5 & M74 & H & LIRIS/WHT & 2013-01-29 & 3860 & 0.5 & $30.0 \pm 0.4^5$ & 0.64 \\
6 & M74 & Ks & LIRIS/WHT & 2012-01-04 & 9105 & 0.7 & $30.0 \pm 0.4^5$ & 0.14 \\
7 & NGC 855 & Ks & LIRIS/WHT & 2012-01-05 & 4350 & 0.8 & $29.94 \pm 0.17^{18}$ & 0.33 \\
8 & NGC 925 & Ks & LIRIS/WHT & 2012-01-01 & 9285 & 0.7 & $29.3 \pm 0.4^9$ & 0.17 \\
9 & NGC 925 & H & LIRIS/WHT & 2013-01-29 & 3000 & 0.6 & $29.3 \pm 0.4^9$ & 0.28 \\
10 & NGC 1058 & Ks & LIRIS/WHT & 2012-01-05 & 5625 & 0.6 & $29.8 \pm 0.4^7$ & 0.24 \\
11 & NGC 1058 & H & LIRIS/WHT & 2013-01-29 & 3000 & 1.0 & $29.8 \pm 0.4^7$ & 0.27 \\
12 & NGC 1313 & Ks & ISAAC/VLT & 2012-07-05 & 160 & 0.6 & $28.2 \pm 0.2^8$ & 0.42 \\
13 & IC342 & Ks & LIRIS/WHT & 2012-01-04 & 7410 & 0.8 & $27.72 \pm 0.17^2$ & 0.21 \\
14 & IC342 & Ks & LIRIS/WHT & 2012-01-01 & 3975 & 1.7 & $27.72 \pm 0.17^2$ & 0.22 \\
15 & IC342 & H & LIRIS/WHT & 2013-01-27 & 4000 & 0.9 & $27.72 \pm 0.17^2$ & 0.23 \\
16 & NGC 1637 & Ks & ISAAC/VLT & 2012-11-22 & 750 & 0.3 & $30.0 \pm 0.4^9$ & 0.4 \\ 
17 & NGC 2403 & Ks & LIRIS/WHT & 2012-01-01 & 6345 & 2.3 & $27.50 \pm 0.05^8$ & 0.28 \\
18 & NGC 2403 & H & LIRIS/WHT & 2013-01-26 & 4000 & 0.7 & $27.50 \pm 0.05^8$ & 0.29 \\
19 & NGC 2403 & Ks & LIRIS/WHT & 2012-01-04 & 7500 & 1.6 & $27.50 \pm 0.05^8$ & 0.28 \\
20 & NGC 2500 & Ks & LIRIS/WHT & 2012-01-01 & 5580 & 1.5 & $30.0 \pm 0.4^7$ & 0.06 \\
21 & NGC 2500 & Ks & LIRIS/WHT & 2012-01-07 & 6855 & 0.6 & $30.0 \pm 0.4^7$ & 0.12 \\
22 & Holmberg II & Ks & LIRIS/WHT & 2012-01-05 & 3750 & 0.9 & $27.65 \pm 0.03^1$ & 0.22 \\
23 & Holmberg II & Ks & LIRIS/WHT & 2013-01-27 & 3675 & 1.0 & $27.65 \pm 0.03^1$ & 0.39 \\
24 & Holmberg II & H & SWIRC/MMT & 2011-05-17 & 1120 & 0.5 & $27.65 \pm 0.03^1$ & 0.23 \\
25 & NGC 2903 & H & LIRIS/WHT & 2013-01-29 & 3940 & 1.0 & $30.1 \pm 0.4^{10}$ & 0.13 \\
26 & Holmberg I& H & SWIRC/MMT & 2011-05-17 & 900 & 0.8 & $27.95 \pm 0.03^1$ & 0.14 \\
27 & Holmberg I & H & SWIRC/MMT & 2011-05-17 & 900 & 0.8 & $27.95 \pm 0.03^1$ & 0.17 \\
28 & Holmberg I & H & SWIRC/MMT & 2011-05-17 & 600 & 1.0 & $27.95 \pm 0.03^1$ & 0.14 \\
29 & M81 & Ks & LIRIS/WHT & 2012-01-08 & 3375 & 0.9 & $27.86 \pm 0.06^6$ & 0.12 \\
30 & Holmberg IX & H & SWIRC/MMT & 2011-05-17 & 1200 & 1.0 & $27.79 \pm 0.08^1$ & 0.25 \\
31 & NGC 3184 & Ks & LIRIS/WHT & 2012-01-07 & 3780 & 1.0 & $30.5 \pm 0.5^5$ & 0.47$^{\dag}$ \\
32 & NGC 3239 & Ks & LIRIS/WHT & 2012-01-01 & 2085 & 1.7 & $29.5 \pm 0.4^7$ & 0.55 \\
33 & NGC 3239 & H & LIRIS/WHT & 2013-01-25 & 3820 & 0.7 & $29.5 \pm 0.4^7$ & 0.39 \\
34 & NGC 3486 & Ks & LIRIS/WHT & 2012-01-07 & 1695 & 0.7 & $31.1 \pm 0.4^{11}$ & 0.30 \\
35 & NGC 3521 & Ks & ISAAC/VLT & 2012-05-16 & 120 & 0.7 & $31.1 \pm 0.4^{10}$ & 0.5$^{\S}$ \\
36 & NGC 3621 & Ks & ISAAC/VLT & 2013-01-01 & 750 & 0.4 & $28.9 \pm 0.4^9$ & 0.23 \\
37 & NGC 3623 & Ks & ISAAC/VLT & 2012-06-09 & 120 & 0.6 & $30.5 \pm 0.4^{10}$ & 0.14 \\
38 & NGC 3627 & Ks & ISAAC/VLT & 2012-06-10 & 120 & 0.7 & $29.7 \pm 0.4^9$ & 0.24 \\
39 & NGC 3628 & Ks & ISAAC/VLT & 2012-06-10 & 120 & 0.9 & $30.7 \pm 0.5^{12}$ & 0.50$^{\dag}$ \\
40 & NGC 4136 & H & LIRIS/WHT & 2013-01-25 & 3980 & 0.9 & $29.9 \pm 0.4^7$ & 0.34 \\
41 & NGC 4204 & H & LIRIS/WHT & 2013-01-29 & 2940 & 1.0 & $29.5 \pm 0.4^7$ & 0.62$^{\dag}$ \\
42 & NGC 4258 & H & LIRIS/WHT & 2013-01-27 & 5000 & 0.8 & $29.29 \pm 0.02^{13}$ & 0.5$^{\S}$ \\
43 & NGC 4258 & H & LIRIS/WHT & 2013-01-27 & 4000 & 0.7 & $29.29 \pm 0.02^{13}$ & 0.31 \\
44 & NGC 4258 & H & LIRIS/WHT & 2013-01-27 & 4000 & 0.7 & $29.29 \pm 0.02^{13}$ & 0.27 \\
45 & NGC 4395 & Ks & LIRIS/WHT & 2012-01-04 & 5595 & 1.2 & $28.42 \pm 0.02^8$ & 0.32 \\
46 & NGC 4449 & Ks & LIRIS/WHT & 2012-01-04 & 3435 & 0.9 & $28.0 \pm 0.1^{14}$ & 0.35 \\
47 & NGC 4559 & H & LIRIS/WHT & 2013-01-25 & 3400 & 0.8 & $29.6 \pm 0.5^{10}$ & 0.28 \\
48 & NGC 4559 & H & SWIRC/MMT & 2011-05-17 & 6680 & 0.8 & $29.6 \pm 0.5^{10}$ & 0.12 \\
49 & NGC 4618 & Ks & LIRIS/WHT & 2012-01-07 & 5265 & 0.9 & $29.3 \pm 0.4^7$ & 0.62 \\
50 & NGC 5128 & Ks & ISAAC/VLT & 2012-04-22 & 120 & 0.3 & $27.77 \pm 0.17^{15}$ & 0.16 \\
51 & NGC 5204 & Ks & LIRIS/WHT & 2011-05-16 & 5040 & 1.0 & $28.3 \pm 0.3^{16}$ & 0.10 \\
52 & M51 & H & LIRIS/WHT & 2013-01-29 & 3460 & 1.2 & $29.6 \pm 0.2^4$ & 0.28 \\ 
53 & M51 & H & LIRIS/WHT & 2013-01-29 & 3140 & 1.3 & $29.6 \pm 0.2^4$ & 0.35 \\
54 & M51 & H & LIRIS/WHT & 2013-01-29 & 3500 & 1.5 & $29.6 \pm 0.2^4$ & 0.23 \\
55 & NGC 5408 & Ks & ISAAC/VLT & 2012-04-22 & 120 & 0.3 & $28.4 \pm 0.8^7$ & 0.26 \\
56 & NGC 5474 & Ks & LIRIS/WHT & 2012-01-05 & 4935 & 1.2 & $29.15^{17}$ & 0.46$^{\dag}$ \\
\hline
   \end{tabular}
  \end{minipage}
 \end{table*}
 
 \begin{table*}
  \begin{minipage}{150mm}
  \contcaption{}
  \begin{tabular}{llccccccc}
  \hline
Im. & Galaxy & Filter & Instrument/ & Obs. date & Exp. time & I.Q. & Distance modulus & WCS uncertainty$^{\ddag}$\\
no. & & & Telescope & & (seconds) & (arcsec) & (magnitudes) & (arcsec) \\
   \hline
57 & M101 & H & SWIRC/MMT & 2011-05-17 & 1160 & 1.2 & $29.04 \pm 0.05^3$ & 0.36 \\
58 & M101 & H & SWIRC/MMT & 2011-05-17 & 1200 & 0.7 & $29.04 \pm 0.05^3$ & 0.23 \\
59 & M101 & H & SWIRC/MMT & 2011-05-17 & 1200 & 0.7 & $29.04 \pm 0.05^3$ & 0.15 \\
60 & M101 & H & SWIRC/MMT & 2011-05-17 & 1180 & 0.9 & $29.04 \pm 0.05^3$ & 0.12 \\
61 & NGC 5585 & Ks & LIRIS/WHT & 2012-01-05 & 4950 & 1.1 & $29.9 \pm 0.6^{10}$ & 0.06 \\
\hline
  \end{tabular}
Notes - $^{\ddag}$: Rms error with respect to the reference catalog. The 2MASS catalog was used for the astrometric calibration, unless stated otherwise. $^*$: Date of the first observation. The second observations were taken on 2012-12-03 (\#1) and 2012-12-31 (\#2). $^{\dag}$: USNO B1.0 catalog used for astrometric calibration. $^{\S}$: less than 4 sources available for astrometric calibration. $^1$: \citet{2009ApJS..183...67D}, $^2$: \citet{2008ApJ...683..630H}, $^3$: \citet{2011ApJ...733..124S}, $^4$: \citet{2009ApJ...694.1067P}, $^5$: \citet{2010ApJ...715..833O}, $^6$: \citet{2010ApJ...718.1118D}, $^7$: \citet{1988ngc..book.....T}, $^8$: \citet{2009AJ....138..332J}, $^9$: \citet{2009AJ....138..323T}, $^{10}$: \citet{2009ApJS..182..474S}, $^{11}$: \citet{2007A&A...465...71T}, $^{12}$: \citet{1997ApJS..109..333W}, $^{13}$: \citet{2008ApJ...672..800H}, $^{14}$: \citet{2010ApJ...721..297M}, $^{15}$: \citet{2009ApJ...705.1533C}, $^{16}$: \citet{2003A&A...398..467K}, $^{17}$: \citet{2000A&AS..142..425D}, $^{18}$: \citet{2001ApJ...546..681T}
 \end{minipage}
\end{table*}

\section{Data reduction and analysis}
The data obtained with VLT/ISAAC in period 89 were reduced with the ISAAC pipeline in the {\sc Gasgano} environment, following the steps outlined in the ISAAC Data Reduction Guide\footnote{ftp.eso.org/pub/dfs/pipelines/isaac/isaac-pipeline-manual-1.4.pdf}. The output of the pipeline is a co-added, sky-subtracted image without astrometric calibration. We used the astrometry tool `fit to star positions' in the {\sc Starlink} program {\sc Gaia} to calculate the astrometric solutions, using 2 Micron All Sky Survey (2MASS, \citealt{skrutskie06}) sources in the field of view if at least four were present. When this was not the case we used sources from the USNO B1.0 catalog \citep{monet03} for the calibration. We adopt the root-mean-square (rms) error of the fit as the 1-$\sigma$ uncertainty on the positions of the NIR sources. 
We used the general data reduction software package {\sc Theli} \citep{schirmer13} to reduce the data obtained with WHT/LIRIS, MMT/SWIRC and the data obtained with VLT/ISAAC in p90. The greater flexibility of {\sc Theli} makes it easier to immediately combine data taken on different nights, hence we decided to use it instead of the ISAAC pipeline for the reduction of the VLT/ISAAC data from p90. Because the final astrometric and photometric calibration was done in a uniform way for all images, we are confident that the use of different data reduction packages does not impact our results. Using {\sc Theli}, we first produced master flats, applied them to the images and subtracted the sky background of the individual frames. {\sc Theli} employs {\sc sExtractor} \citep{bertin96} and {\sc Scamp} \citep{bertin06} to find sources in the individual images and determine an astrometric solution by matching their positions to sources from 2MASS or PPXML \citep{roeser10} if not enough 2MASS sources are present. This astrometric solution is then used to coadd the images using {\sc Swarp} \citep{bertin02}. We improved the astrometric accuracy of the final coadded images using the `fit to star positions' tool in {\sc Gaia}, again fitting to positions of sources from 2MASS or USNO B1.0. The rms errors on these fits are listed in Table \ref{logtabel} as a measure of the uncertainty in the astrometry of the NIR images. 
Three images (\#20, 45 and 52) did not have a sufficient number of 2MASS or USNO B1.0 sources in the field of view. We calibrated these images using only three reference stars from 2MASS. With three reference stars and three degrees of freedom no rms error could be calculated. For these images we adopt a 1-$\sigma$ uncertainty of 0.5 arcsec.

We calculated the radius of the 95\% confidence (2-$\sigma$) error circle around the position of the ULX candidates on the NIR images by quadratically adding the 2-$\sigma$ uncertainty on the X-ray position and twice the rms error of the astrometric solution of the NIR images. There is an additional uncertainty in the 2MASS and USNO B1.0 positions with respect to the international celestial reference system (ICRS), but this error (0.015 arcsec for 2MASS, 0.2 arcsec for USNO B1.0, \citealt{skrutskie06,monet03}) is negligible in the total error budget (that is dominated by the uncertainty in the X-ray position).

We used the tasks DAOFIND and PHOT from the {\sc Apphot} package in {\sc Iraf}\footnote{{\sc IRAF} is distributed by the National Optical Astronomy Observatory, which is operated by the Association of Universities for Research in Astronomy (AURA) under cooperative agreement with the National Science Foundation.} to calibrate the zeropoints of the images using 2MASS sources near the targets, detect counterparts within the error circles and calculate their apparent magnitude. For the cases where we did not detect a counterpart we calculated the limiting magnitude at the position of the ULX by simulating stars at that position with a range of magnitudes using the {\sc Iraf} task MKOBJECTS. The magnitude of the faintest simulated star that was still detected at the three sigma level was taken as a robust lower limit for the magnitude of the ULX.

\section{Results}
Of the 62 ULX candidates in our sample, 17 have a candidate counterpart in one or more NIR images. The apparent and absolute magnitudes of the candidate counterparts are listed in Table \ref{resultstabel}. If a ULX has no detected counterpart, the limiting magnitude of the image at the position of the ULX is given. The errors on the apparent magnitudes reported in Table \ref{resultstabel} are the 1-$\sigma$ errors on the magnitude determination by {\sc Iraf} and on the zeropoint calibration. For the absolute magnitudes there is an additional error that stems from the uncertainty in the distance modulus (listed in Table \ref{logtabel}).

\subsection{NIR counterparts}
We detected candidate counterparts for 17 ULX candidates in our sample. Because some of these have been observed more than once, the total number of detections is 23. There are 11 sources in the \textit{Ks}-band  and 12 in the $H$-band. The absolute magnitudes of the candidate counterparts range from -7.1 to -13.88 in the $H$-band and from -8.1 to -14.3 in the \textit{Ks}-band, with the majority of the sources lying between -9 and -11.5 in both bands (see Table \ref{resultstabel}).

\subsection{Limiting magnitudes for non-detections}
For the remaining 45 ULX candidates in our sample we only have lower limits on their NIR magnitudes, 29 in the \textit{Ks}-band and 23 in the $H$-band. The apparent limiting magnitudes range from 17 to 21.25 in both NIR bands. Our aim was to reach a limiting magnitude of at least 20. Worse limits are mainly caused by bad seeing (although most of the sources that were observed under bad seeing conditions were subsequently repeated under better conditions) and crowding or high backgrounds from the host galaxies, especially in e.g. the spiral arms of M101 and M51. 

The absolute limiting magnitudes vary from from -7.5 to -12 in the \textit{Ks}-band and from -7.5 to -12.6 in the $H$-band.

\begin{table*}
 \begin{minipage}{160mm}
 \caption{Apparent and absolute magnitudes of the candidate NIR counterparts to the ULX candidates, or upper limits for ULXs where we did not detect a source in the error circle around the X-ray position. The image number refers to the observation log (Table \ref{logtabel}). The error radius is the radius of the 95\% confidence circle around the position of the ULX within which we search for counterparts. It is derived by quadratically adding the positional error on the location of the X-ray sources and two times the rms error of the astrometric solution of fitting stellar positions on the NIR images.}\label{resultstabel}
  \begin{tabular}{llcccccc}
  \hline
  Galaxy & ULX name & Im. & Error & Det. & Filter & Apparent magnitude$^{\dagger}$ & Absolute magnitude$^{\dagger}$ \\
  & & no. & Radius & y/n & & & \\
  \hline
NGC 253 & J004722-252051 & 1 & 1.1'' & y & Ks & $17.2 \pm 0.03 \pm 0.5$ & $-10.5 \pm 0.03 \pm 0.5 \pm 0.10$ \\
NGC 253 & J004742-251501 & 2 & 1.4'' & n & Ks & $> 19.5$ & $> -8.2$\\
M74 & M74 X-1 & 3 & 1.2'' & n & Ks & $> 20.0$ & $> -10.0$\\
 & & 4 & 1.0'' & n & Ks & $> 20.0$ & $> -10.0$\\
 & & 5 & 1.5'' & n & H & $> 21.25$ & $> -8.8$\\
M74 & J013636+155036 & 6 & 0.8'' & n & Ks & $> 21.25$ & $> -8.8$\\
NGC 855 & J021404+275239 & 7 & 1.2'' & n & Ks & $> 18.25$ & $> -11.7$\\
NGC 925 & J022721+333500 & 8 & 1.1'' & y & Ks & $18.0 \pm 0.03 \pm 0.2$ & $-11.3 \pm 0.03 \pm 0.2 \pm 0.4$ \\
 & & 9 & 1.1'' & y & H & $18.7 \pm 0.03 \pm 0.2$ & $-10.6 \pm 0.03 \pm 0.2 \pm 0.4$ \\
NGC 925 & J022727+333443 & 8 & 1.1'' & y & Ks & $19.5 \pm 0.08 \pm 0.2$ & $-9.8 \pm 0.08 \pm 0.2 \pm 0.4$ \\
 & & 9 & 1.1'' & y & H & $20.1 \pm 0.08 \pm 0.2$ & $-9.2 \pm 0.08 \pm 0.2 \pm 0.4$ \\
NGC 1058 & J024323+372038 & 10 & 0.8'' & y & Ks & $19.7 \pm 0.06 \pm 0.4$ & $-10.1 \pm 0.06 \pm 0.4 \pm 0.4$ \\
 & & 11 & 0.9'' & y & H & $20.8 \pm 0.2 \pm 0.3$ & $-9.0 \pm 0.2 \pm 0.3 \pm 0.4$\\
NGC 1313 & J0318-6629 & 12 & 1.3'' & n & Ks & $> 18.5$ & $> -9.7$\\
IC342 & J034555+680455 & 13 & 0.5'' & n & Ks & $> 20.25$ & $> -7.5$\\
IC342 & J034615+681112 & 14 & 0.6'' & n & Ks & $> 18.5$ & $> -9.2$\\
 & & 15 & 0.6'' & n & H & $> 20.0$ & $> -7.7$\\
NGC 1637 & [IWL2003] 68 & 16 & 1.3'' & y & Ks & $16.3 \pm 0.005 \pm 0.5$ & $-13.7 \pm 0.005 \pm 0.5 \pm 0.4$ \\
NGC 2403 & J073625+653539 & 17 & 0.9'' & n & Ks & $> 18.25$ & $> -9.3$\\
 & & 18 & 0.9'' & n & H & $> 20.0$ & $> -7.5$ \\
NGC 2403 & J0732+6546 & 19 & 0.9'' & n & Ks & $> 19.25$ & $> -8.3$ \\
NGC 2500 & J080148+504354 & 20 & 1.0'' & n & Ks & $> 19.75$ & $> -10.3$\\
NGC 2500 & J080157+504339 & 20 & 1.0'' & y & Ks & $15.7 \pm 0.002 \pm 0.2$ & $-14.3 \pm 0.002 \pm 0.2 \pm 0.4$ \\
 & & 21 & 1.0'' & y & Ks & $15.95 \pm 0.005 \pm 0.15$ & $-14.1 \pm 0.005 \pm 0.15 \pm 0.4$ \\
Holmberg II & Holmberg II X-1 & 22 & 0.8'' & y & Ks & $19.30 \pm 0.08 \pm 0.10$ & $-8.35 \pm 0.08 \pm 0.10 \pm 0.03$ \\
 & & 23 & 1.0'' & y & Ks & $19.4 \pm 0.12 \pm 0.2$ & $-8.2 \pm 0.12 \pm 0.2 \pm 0.03$ \\
 & & 24 & 0.8'' & y & H & $20.6 \pm 0.3 \pm 0.10$ & $-7.1 \pm 0.3 \pm 0.10 \pm 0.03$ \\
NGC 2903 & J093206+213058 & 25 & 0.7'' & n & H & $> 20.25$ & $> -9.9$ \\
Holmberg I & Ho I XMM1 & 26 & 1.0'' & y & H & $17.81 \pm 0.01 \pm 0.10$ & $-10.14 \pm 0.01 \pm 0.10 \pm 0.03$ \\
Holmberg I & J0940+7106 & 27 & 1.1'' & n & H & $> 19.25$ & $> -8.7$ \\
Holmberg I & Ho I XMM3 & 28 & 1.4'' & n & H & $> 20.5$ & $> -9.6$ \\
M81 & NGC 3031 ULX1 & 29 & 1.0'' & n & Ks & $> 18.5$ & $> -9.4$ \\
Holmberg IX & Holmberg IX X-1 & 30 & 1.2'' & n & H & $> 19.75$ & $> -8.0$ \\
NGC 3184 & J101812+412421 & 31 & 1.2'' & n & Ks & $> 20.5$ & $> -10.0$ \\
NGC 3239 & J102506+170947 & 32 & 1.5'' & n & Ks & $> 18.25$ & $> -11.3$ \\
 & & 33 & 1.3'' & n & H & $> 20.25$ & $> -9.3$ \\
NGC 3239 & J102508+170948 & 32 & 1.5'' & n & Ks & $> 17.5$ & $> -12.0$ \\
 & & 33 & 1.3'' & n & H & $> 20.25$ & $> -9.3$ \\
NGC 3486 & J110022+285818 & 34 & 0.9'' & n & Ks & $> 20.0$ & $> -11.1$ \\
NGC 3521 & J110545+000016 & 35 & 1.4'' & n & Ks & $> 19.25$ & $> -11.9$ \\
NGC 3621 & NGC3621 B & 36 & 0.8'' & n & Ks & $> 18.0$ & $> -10.9$ \\
NGC 3623 & NGC 3623 ULX1 & 37 & 1.0'' & n & Ks & $> 19.75$ & $> -10.8$ \\
NGC 3627 & J112020+125846 & 38 & 1.1'' & n & Ks & $> 20.0$ & $> -9.7$ \\
NGC 3627 & J112018+125900 & 38 & 1.1'' & y & Ks & $20.6 \pm 1.9 \pm 0.7$ & $-9.1 \pm 1.9 \pm 0.7 \pm 0.4$ \\
NGC 3628 & J112037+133429 & 39 & 1.4'' & n & Ks & $> 19.5$ & $> -11.2$ \\
NGC 4136 & J120922+295551 & 40 & 1.2'' & y & H & $19.13 \pm 0.03 \pm 0.10$ & $-10.78 \pm 0.03 \pm 0.10 \pm 0.4$ \\
NGC 4136 & J120922+295559 & 40 & 1.2'' & y & H & $19.15 \pm 0.03 \pm 0.10$ & $-10.75 \pm 0.03 \pm 0.10 \pm 0.4$ \\
NGC 4204 & J121510+203912 & 41 & 1.6'' & n & H & $> 20.25$ & $> -9.3$ \\
NGC 4258 & J121844+471730 & 42 & 1.4'' & n & H & $17.79 \pm 0.02 \pm 0.10$ & $-11.50 \pm 0.02 \pm 0.1 \pm 0.02$ \\
NGC 4258 & J121857+471558 & 43 & 1.2'' & n & H & $> 19.0$ & $> -10.3$ \\
NGC 4258 & NGC4258 XMM1 & 42 & 1.3'' & n & H & $> 19.5$ & $> -9.8$ \\
NGC 4258 & J121845+472420 & 44 & 1.1'' & n & H & $> 21.0$ & $> -8.3$ \\
NGC 4395 & IXO 53 & 45 & 1.2'' & n & Ks & $> 20.0$ & $> -8.4$ \\
NGC 4449 & J122818+440634 & 46 & 1.3'' & n & Ks & $> 18.25$ & $> -9.8$ \\
NGC 4559 & J123551+27561 & 47 & 0.9'' & n & H & $> 21.0$ & $> -8.6$ \\
 & & 48 & 0.7'' & n & H & $> 21.25$ & $> -8.4$ \\
NGC 4618 & J124129+410757 & 49 & 1.4'' & n & Ks & $> 20.25$ & $> -9.1$\\
NGC 5128 & J132519-430312 & 50 & 1.0'' & n & Ks & $> 18.5$ & $> -9.3$ \\
NGC 5128 & J132518-430304 & 50 & 0.5'' & n & Ks & $> 19.25$ & $> -8.5$ \\
NGC 5204 & J132938+582506 & 51 & 1.0'' & n & Ks & $> 19.0$ & $> -9.3$ \\
\hline
  \end{tabular}
 \end{minipage}
\end{table*}

\begin{table*}
 \begin{minipage}{160mm}
 \contcaption{}
  \begin{tabular}{llcccccc}
  \hline
  Galaxy & ULX & Im. & Error & Det. & Filter & Apparent magnitude & Absolute magnitude \\
  & & no. & Radius & y/n & & & \\
  \hline
M51 & J132943+47115 & 52 & 0.9'' & n & H & $> 17.5$ & $> -12.1$ \\
M51 & J132950+471153 & 52 & 0.9'' & n & H & $> 17.25$ & $> -12.4$ \\
M51 & J132953+471040 & 52 & 0.9'' & y & H & $15.72 \pm 0.02 \pm 0.10$ & $-13.88 \pm 0.02 \pm 0.10 \pm 0.2$ \\
M51 & J132954+47145 & 53 & 1.0'' & n & H & $> 20.5$ & $> -9.1$ \\
M51 & J132957+471047 & 52 & 0.9'' & n & H & $> 17.5$ & $> -12.1$ \\
M51 & J133001+47137 & 52 & 0.9'' & n & H & $> 17.0$ & $> -12.6$ \\
M51 & J133006+47156 & 53 & 1.0'' & n & H & $> 18.5$ & $> -11.1$ \\
M51 & J133007+47110 & 54 & 0.8'' & n & H & $> 19.5$ & $> -10.1$ \\
NGC 5408 & NGC 5408 X-1 & 55 & 1.3'' & y & Ks & $20.3 \pm 0.13 \pm 0.2$ & $-8.1 \pm 0.13 \pm 0.2 \pm 0.8$ \\
NGC 5474 & NGC 5474-X1 & 56 & 1.4'' & n & Ks & $> 19.5$ & $> -9.7$ \\
M101 & J140332+542103 & 57 & 1.0'' & n & H & $> 18.5$ & $> -10.5$ \\
M101 & J1402+5440 & 58 & 0.8'' & y & H & $19.3 \pm 0.04 \pm 0.2$ & $-9.7 \pm 0.04 \pm 0.2 \pm 0.05$ \\
M101 & J140248+541350 & 59 & 0.8'' & y & H & $17.72 \pm 0.01 \pm 0.05$ & $-11.32 \pm 0.01 \pm 0.05 \pm 0.05$ \\
M101 & J140314+541807 & 60 & 0.7'' & y & H & $18.35 \pm 0.03 \pm 0.10$ & $-10.69 \pm 0.03 \pm 0.10 \pm 0.05$ \\
NGC 5585 & J141939+564137 & 61 & 1.0'' & n & Ks & $> 19.75$ & $> -10.2$ \\
\hline
\end{tabular}
Notes - $^{\dagger}$: Magnitude and errors are as follows: Apparent/absolute magnitude $\pm$ statistical error on the magnitude as given by the PHOT routine $\pm$ statistical error on the zeropoint of the image $\pm$ error on the distance modulus (absolute magnitude only). 
 \end{minipage}
\end{table*}

\section{Discussion}
We obtained NIR ($H$- and \textit{Ks}-band) images of 62 ULXs in 37 nearby galaxies to search for NIR counterparts that could be red supergiant companions. For 17 ULXs, we detected a candidate counterpart in one or more images. Twelve of these have absolute magnitudes that are consistent with those of red supergiants (whose absolute magnitudes range from -8 (K0) to -10.5 (M5) in the $K$-band, \citealt{elias85,drilling00}). One of these, Ho I XMM1, we suspect to be a background active galactic nucleus (AGN; see Section 6.2); we can not exclude the possibility that other RSG candidates are also background AGN. The remaining five sources have an absolute magnitude $< -11$. These brightest sources are not likely to be single stars. They could be background AGN or (young) stellar clusters; see Section 6.2 for more details on the individual sources. 

For the other 45 sources in our sample we obtained limiting magnitudes. Several of these ULX candidates have known optical counterparts, although in many cases it is not clear what is the source of the optical emission: a companion star or emission from an irradiated accretion disc.
 
\begin{figure}
\includegraphics[width=0.5\textwidth]{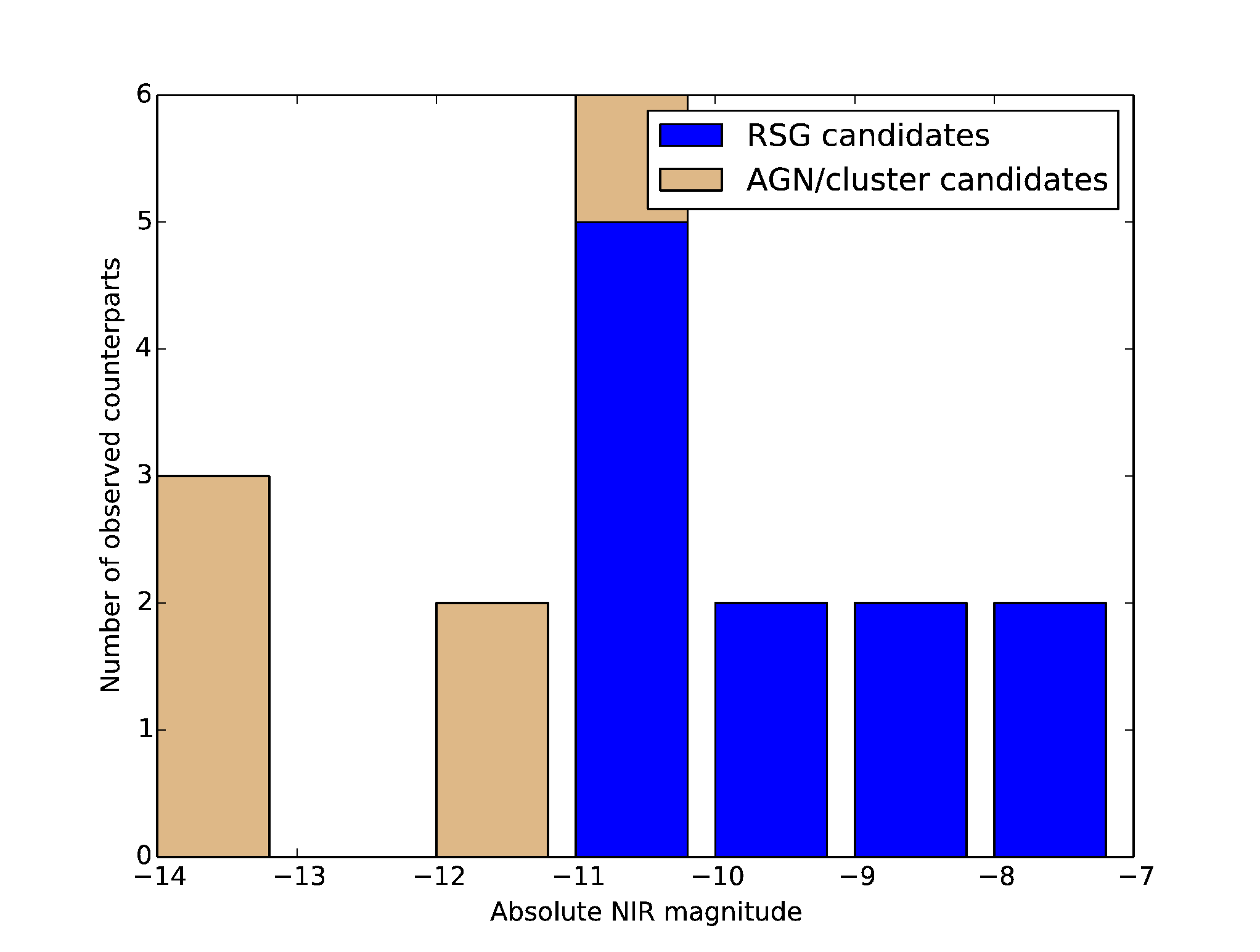}
\caption{The distribution of absolute magnitudes of the detected candidate counterparts. For ULXs with both $H$- and \textit{Ks}-band counterparts only the $H$-band magnitude is included.}\label{hist}
\end{figure}

The distribution of absolute magnitudes of the detected counterparts is plotted in Figure \ref{hist}. At the faint end our sample is not complete, so nothing can be inferred from the shape of the distribution. At absolute magnitudes below -11, however, we expect to detect nearly all NIR sources. This is also the magnitude that the brightest single red supergiants can reach. The sharp drop in the number of detections below that magnitude therefore seems to imply that we do detect a population of possible single stars, the more so because the brightest sources are either confirmed background AGN or strong candidates to be background AGN or star clusters.

The fact that only a fraction of the ULXs has a bright NIR counterpart could reflect a difference between systems with blue supergiant or Wolf-Rayet donor stars and systems with red supergiant donor stars. The ratio of blue to red supergiants is not well known, and the ratio of non-detections to candidate RSG detections in our sample of $45/11 \approx 4.1$ is consistent with what has been observed \citep{langer95}. Alternatively, the systems where we detected NIR counterparts might have larger accretion discs than the ones without a NIR detection, causing the excess NIR radiation. This scenario would suggest that the black holes in those ULXs have larger Roche lobes, either because they have a higher mass than the black holes in systems without bright NIR counterparts (assuming they have the same companion stars), or because the initial separation between the black hole and companion star was larger and hence, when the contact phase starts, the companion star is more evolved and both the disc 
and star are larger.

Another possibility is that the infrared radiation originates from a jet. We investigated this scenario for the only ULX in our sample with a NIR counterpart and an observed radio jet: Holmberg II X-1. We find that the counterpart that we detected is $\sim 6$ magnitudes brighter than what we would expect as contribution from the jet if it has a steep spectrum ($\alpha \approx -0.8$, for $S_{\nu} \propto \nu^{\alpha}$, as is common for jets from X-ray binaries accreting at high rates; \citealt{fender09}). If the radio emission is due to a steady jet with a flat spectrum ($\alpha = 0$) it would contribute significantly to the NIR emission, but such jets usually appear in the low/hard state (see Section 6.1 - Holmberg II X-1). Finding a ULX in the low/hard state would be strong evidence for an IMBH. In principle the X-ray spectra of ULXs in the `ultraluminous' state (showing a soft excess and a rollover in the spectrum above 3 keV, \citealt{gladstone09}) can be interpreted as a low/hard state with X-ray reflection from an ionized disc \citep{caballerogarcia10}. However, this interpretation has been ruled out at least for a number of ULXs (e.g. \citealt{bachetti13}) by observations of their hard X-ray spectra with \textit{NuSTAR} \citep{harrison13}, and thus far only HLX-1 has been observed in a state that is reminiscent of the low/hard state \citep{godet09}.

In Sections 6.1 and 6.2 we discuss the systems with NIR counterparts in more detail.

\begin{figure*}
 \hbox{
 \includegraphics[width=0.33\textwidth]{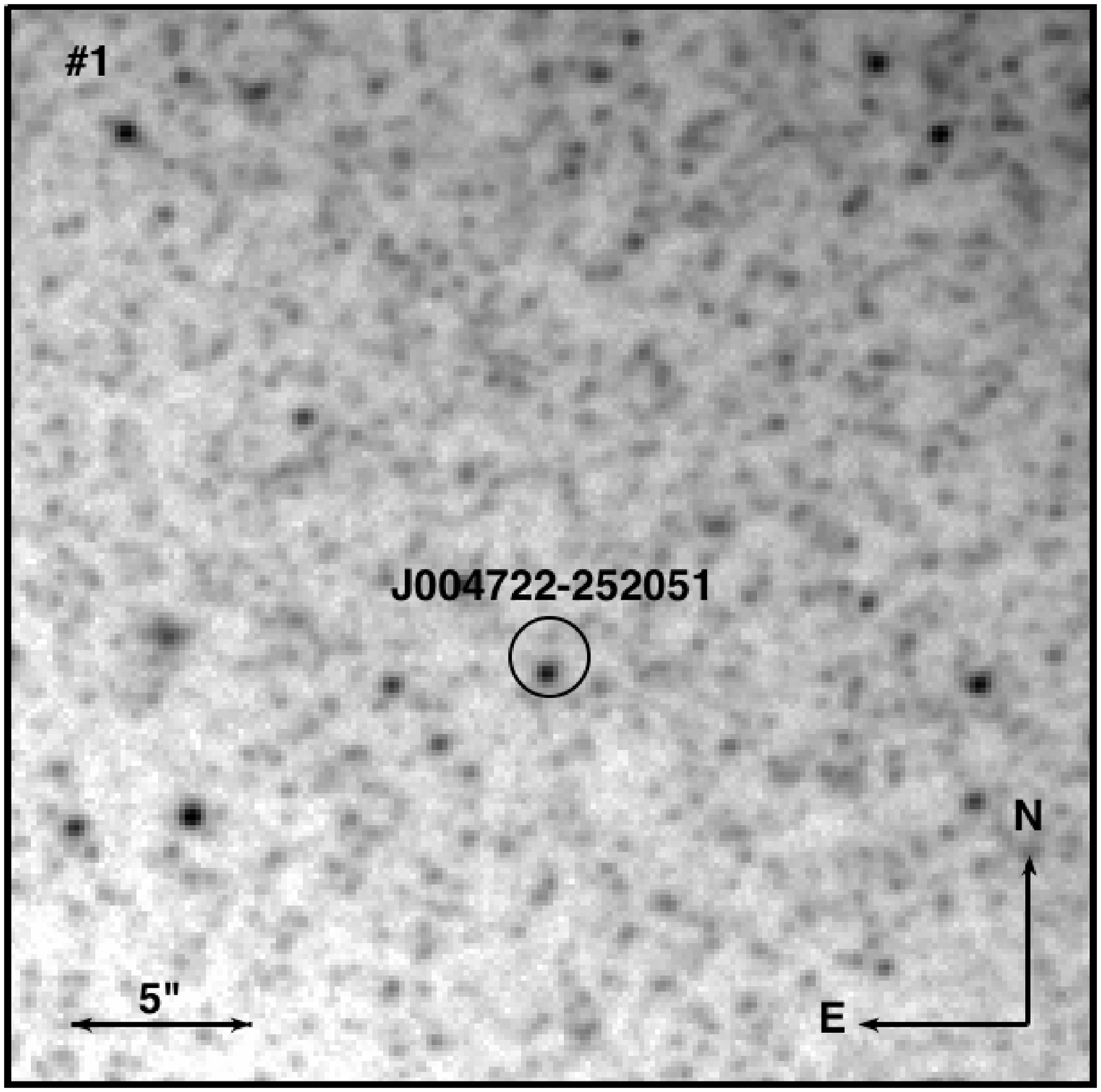}
 \includegraphics[width=0.33\textwidth]{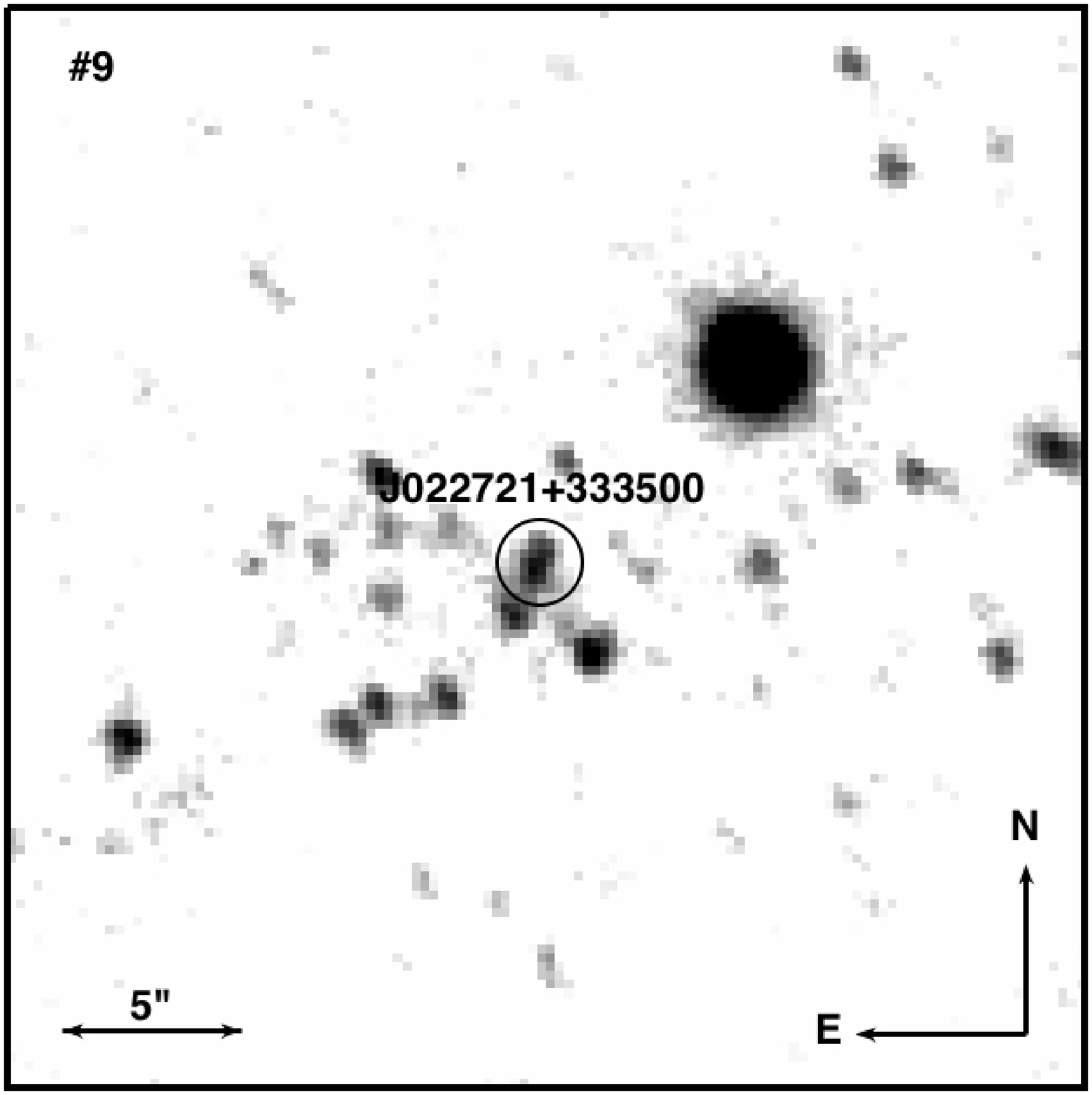}
 \includegraphics[width=0.33\textwidth]{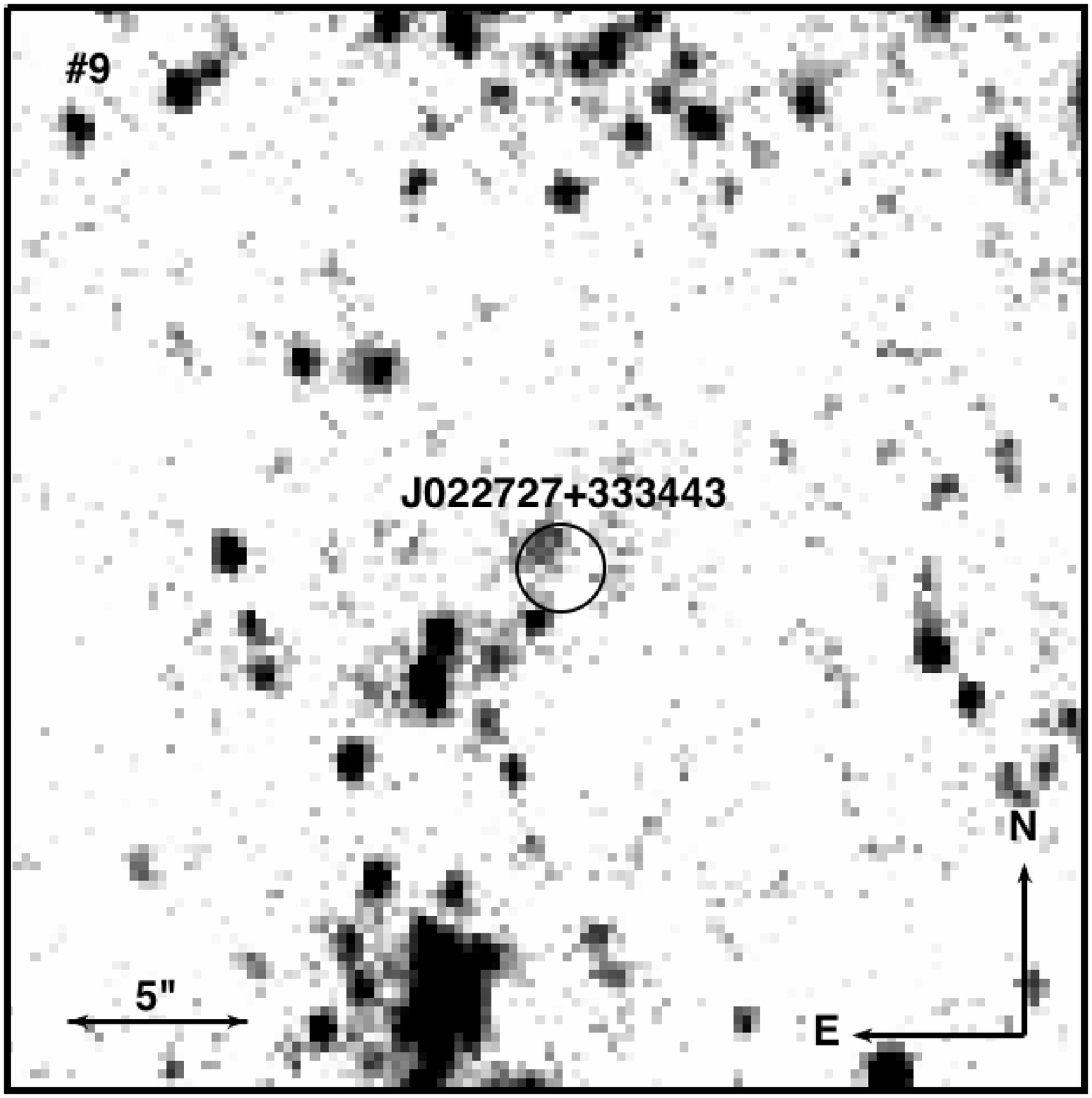}
 }
 \hbox{
 \includegraphics[width=0.33\textwidth]{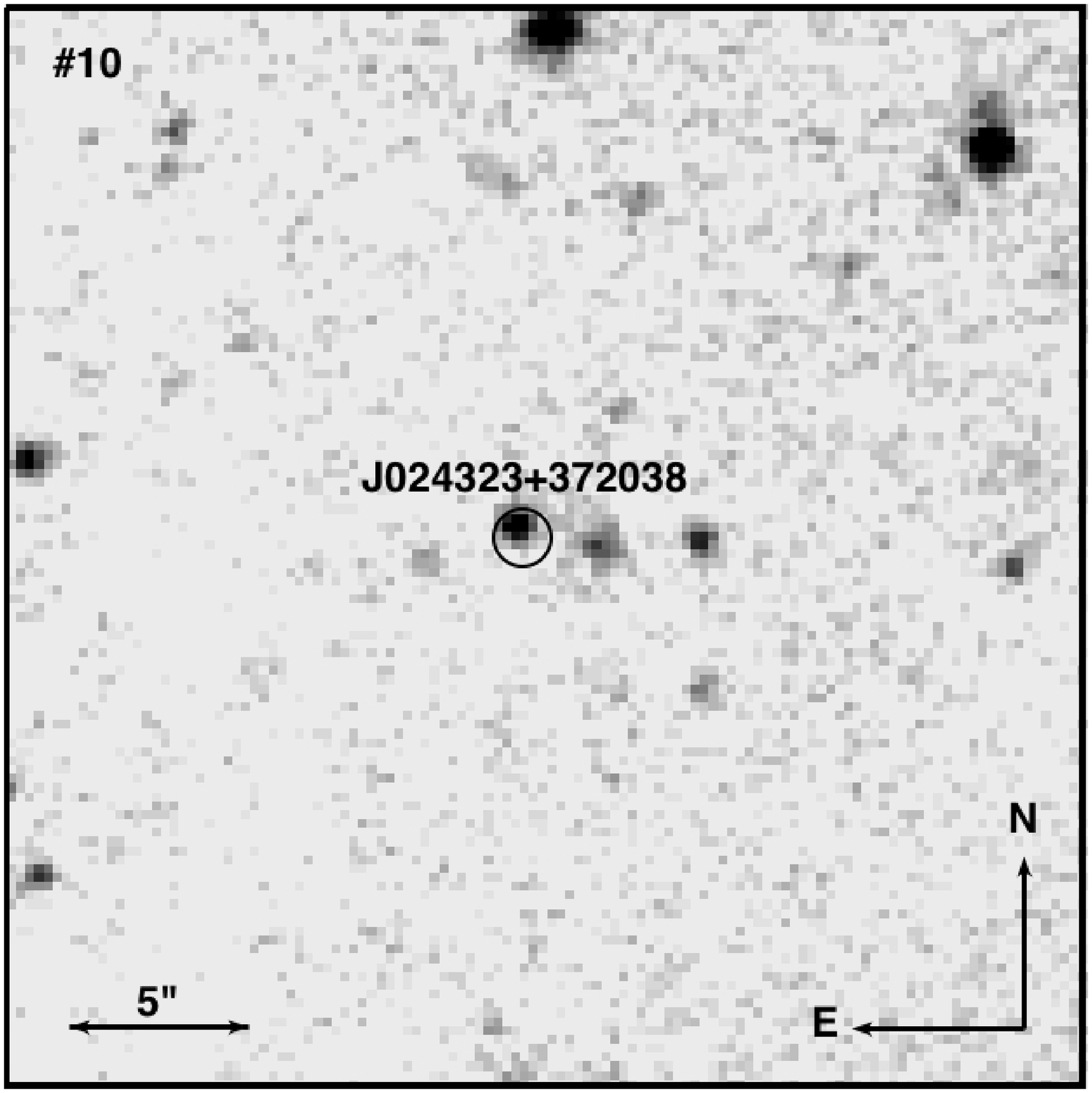}
 \includegraphics[width=0.33\textwidth]{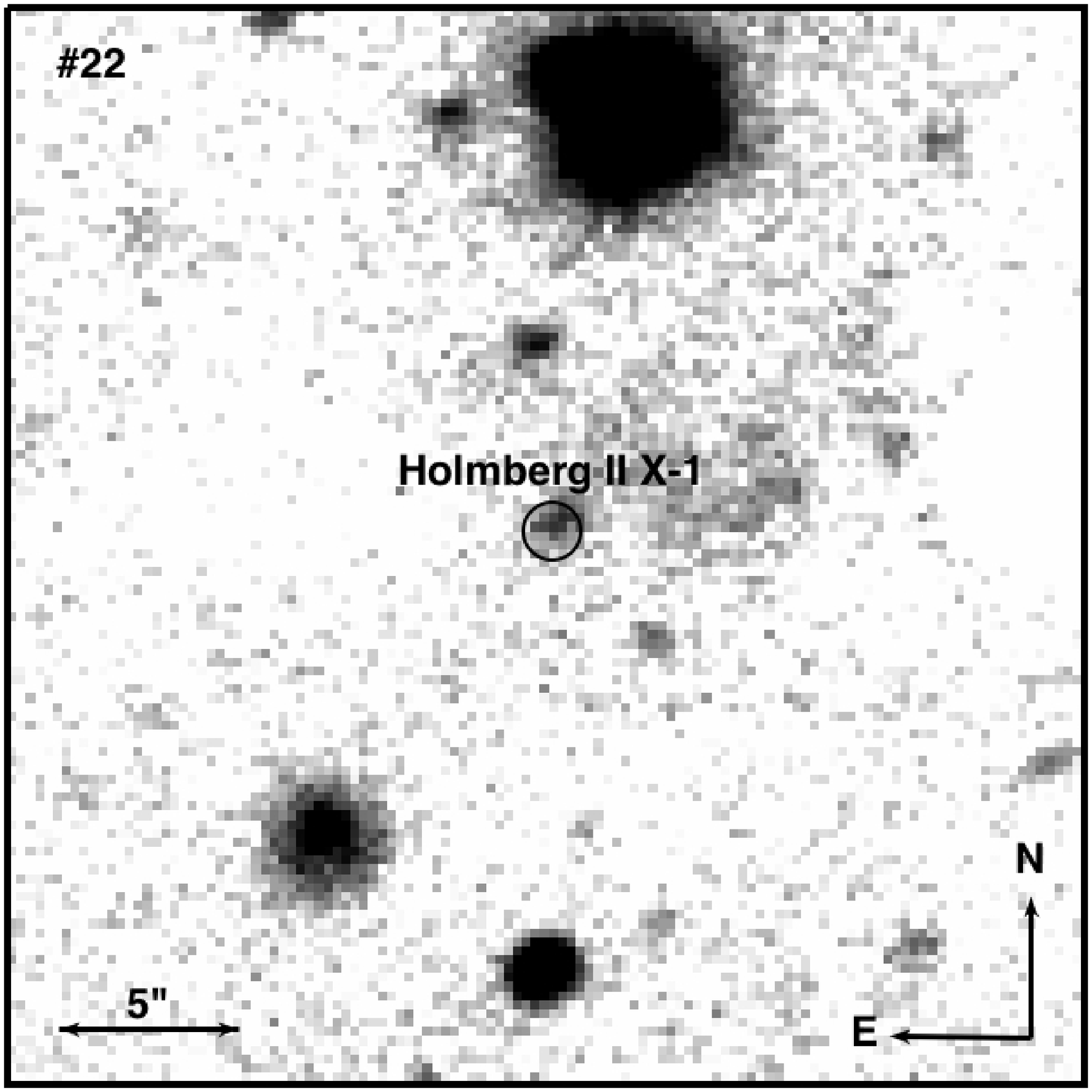}
 \includegraphics[width=0.33\textwidth]{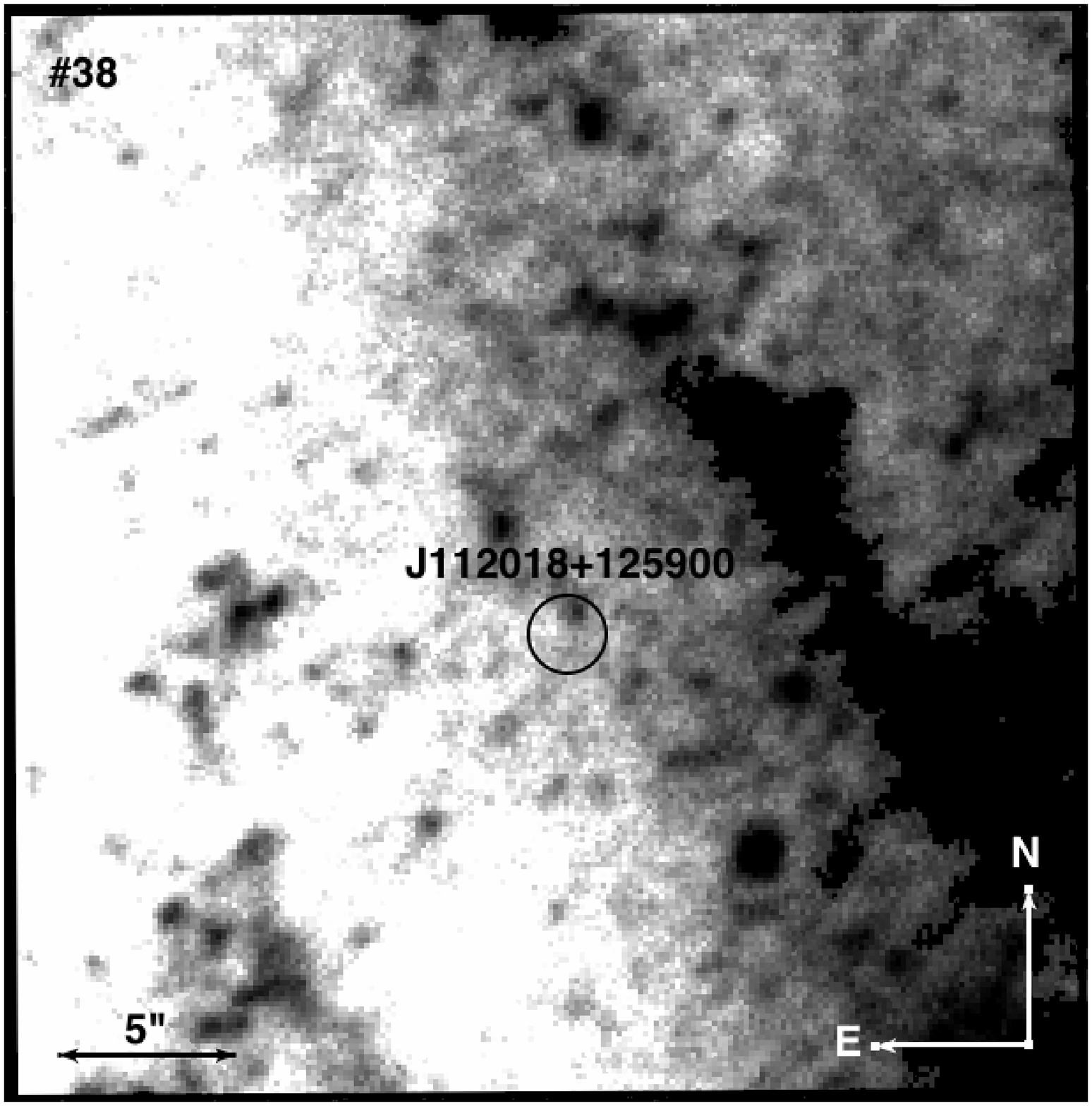}
}
 \hbox{
 \includegraphics[width=0.33\textwidth]{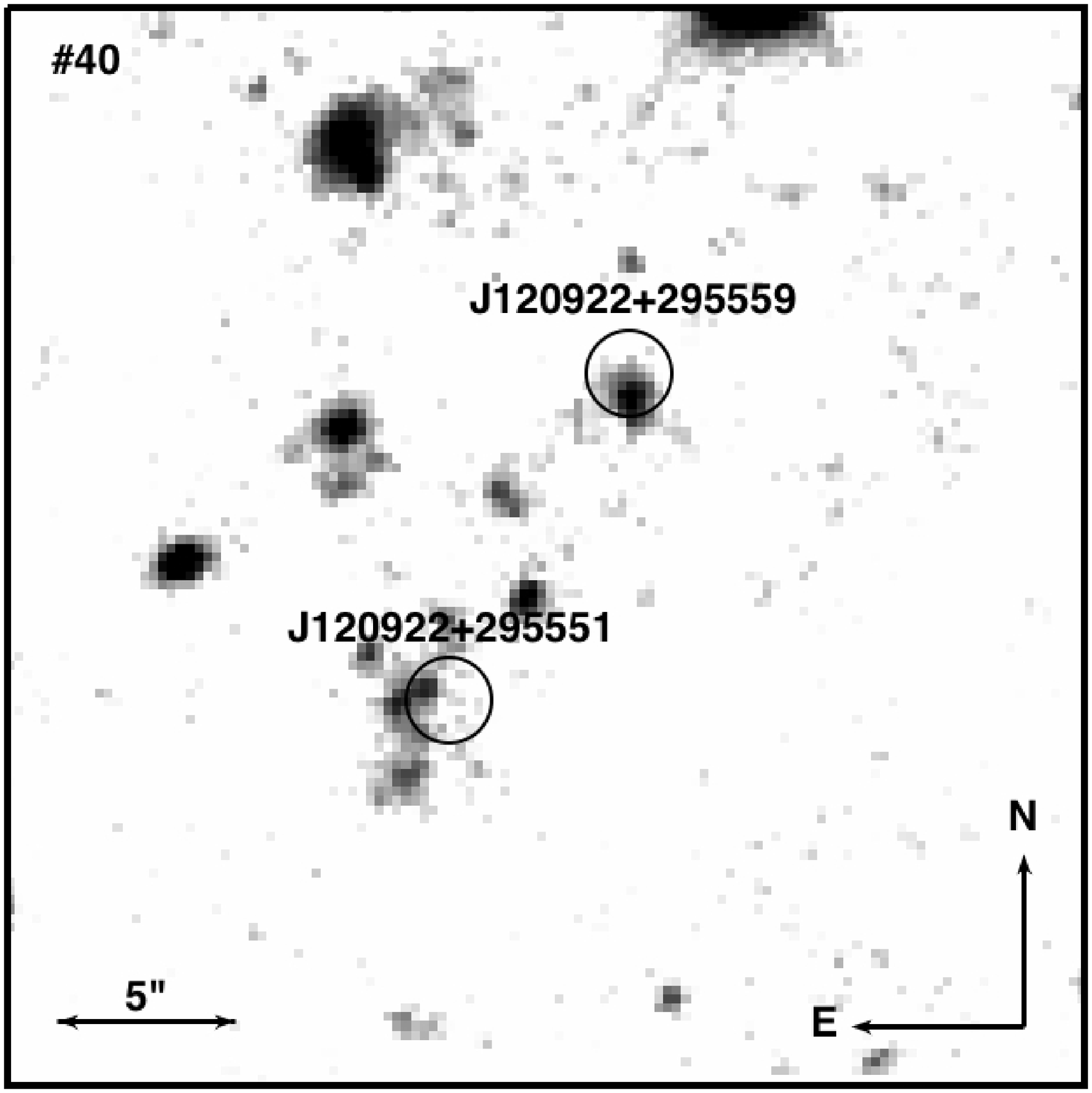}
 \includegraphics[width=0.33\textwidth]{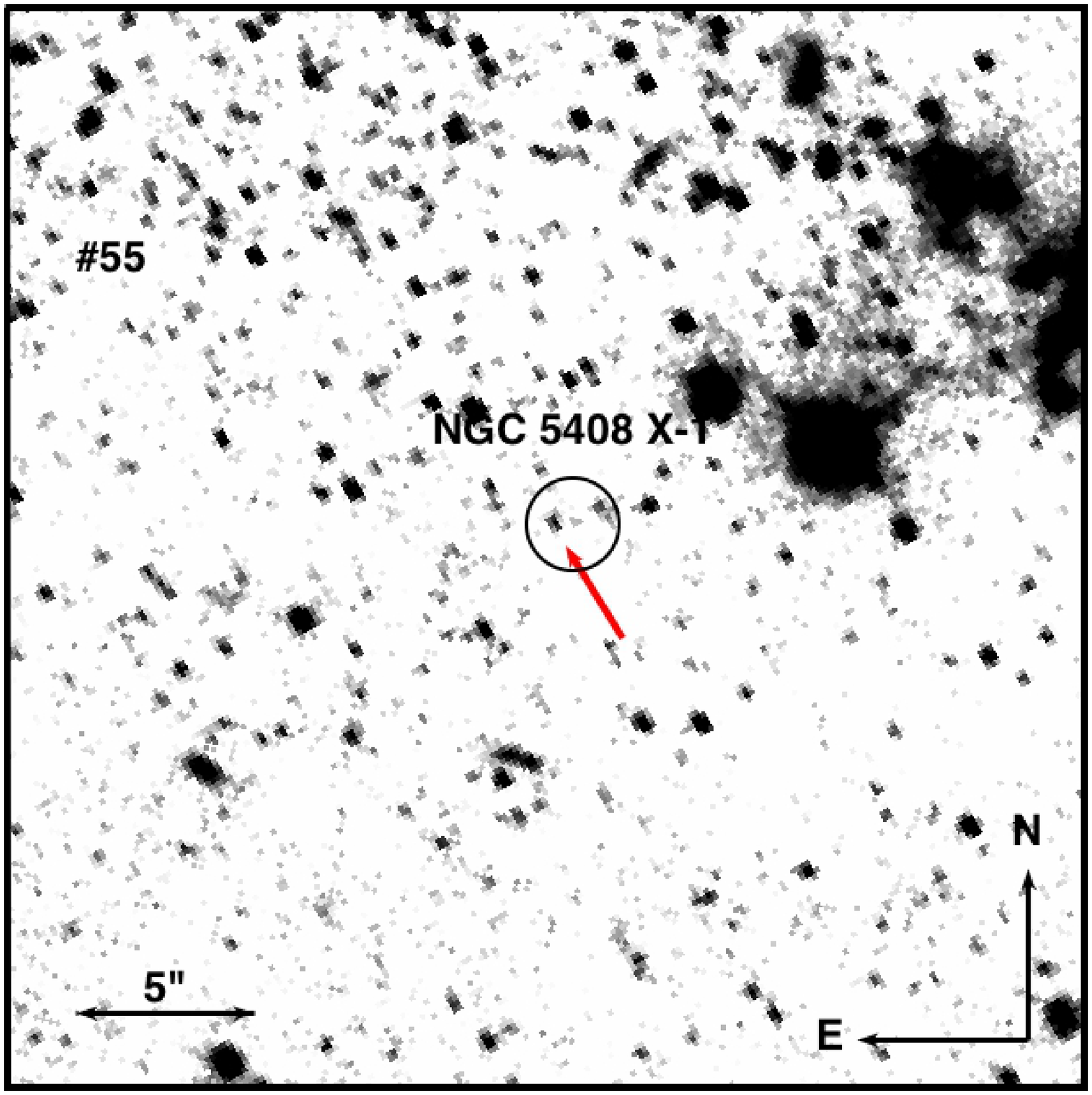}
 \includegraphics[width=0.33\textwidth]{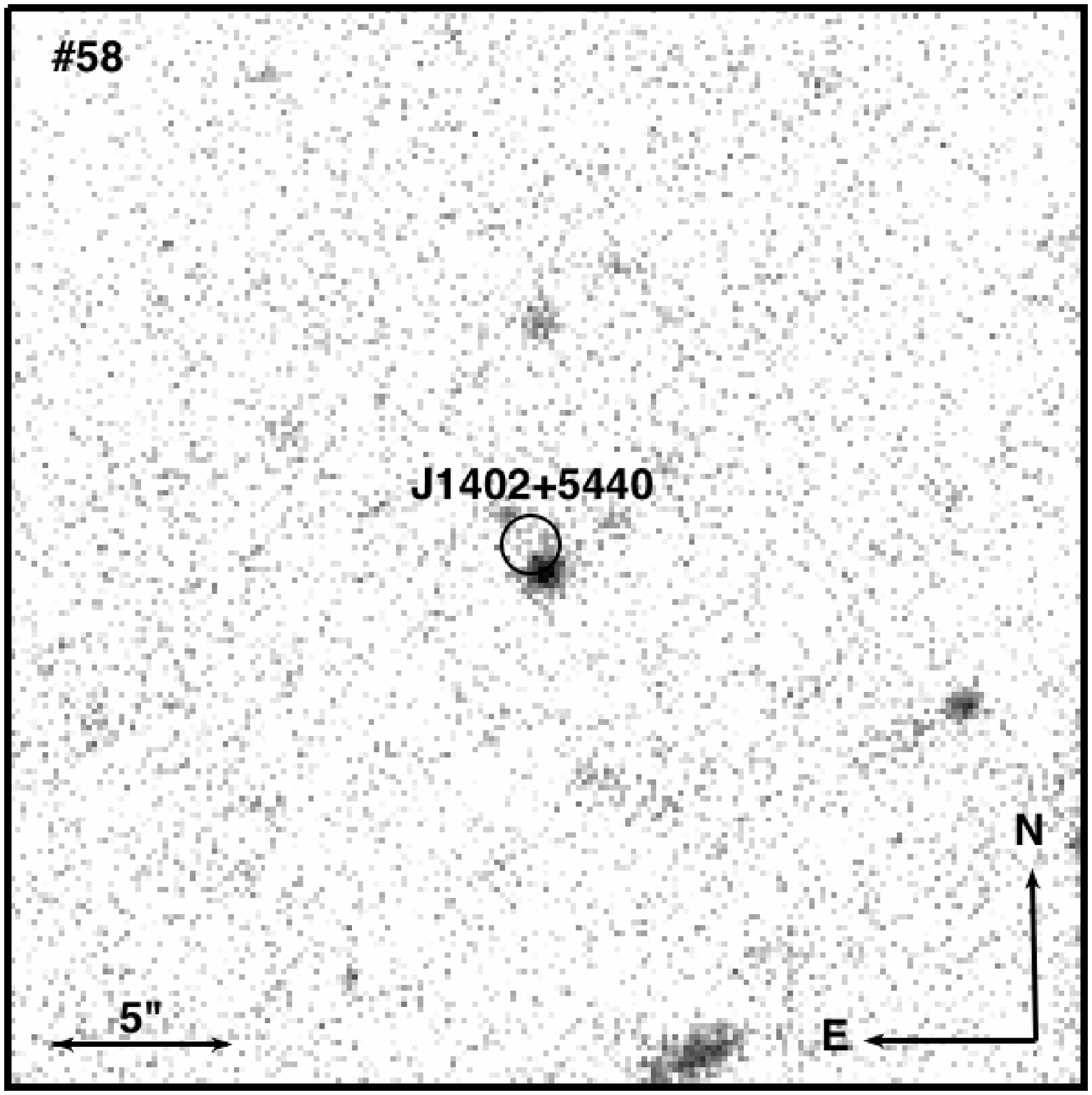}
}
 \caption{0.5' $\times$ 0.5' finder charts of all ULXs with a candidate counterpart that could be a RSG. The black circles are the 95\% confidence error circles around the positions of the ULXs. The numbers in the upper left corners refer to the image numbers in Table \ref{logtabel}.}\label{rsgfinders}
\end{figure*}

\begin{figure*}
 \hbox{
  \includegraphics[width=0.33\textwidth]{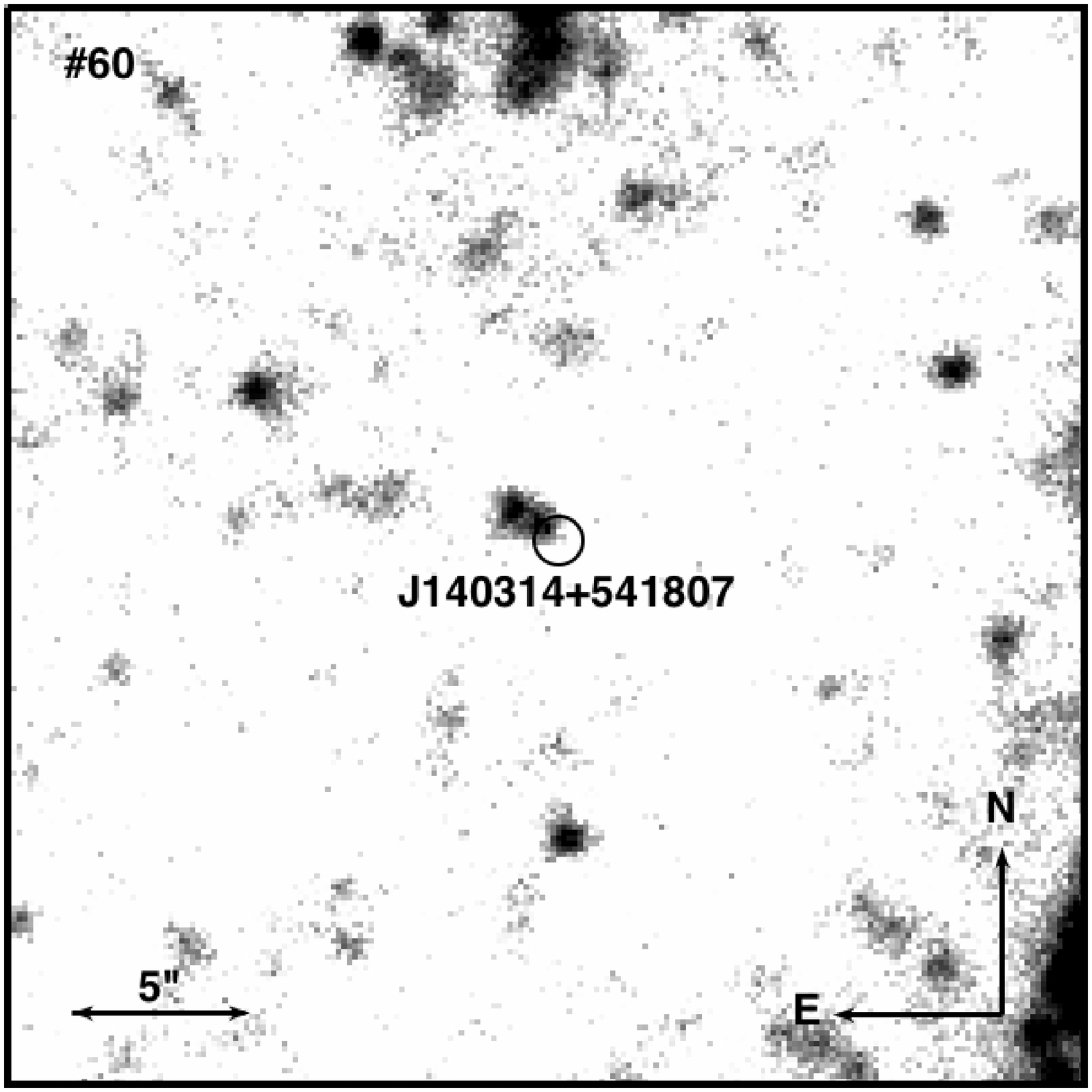}
 }
 \contcaption{}
\end{figure*}

\begin{figure*}
 \hbox{
 \includegraphics[width=0.33\textwidth]{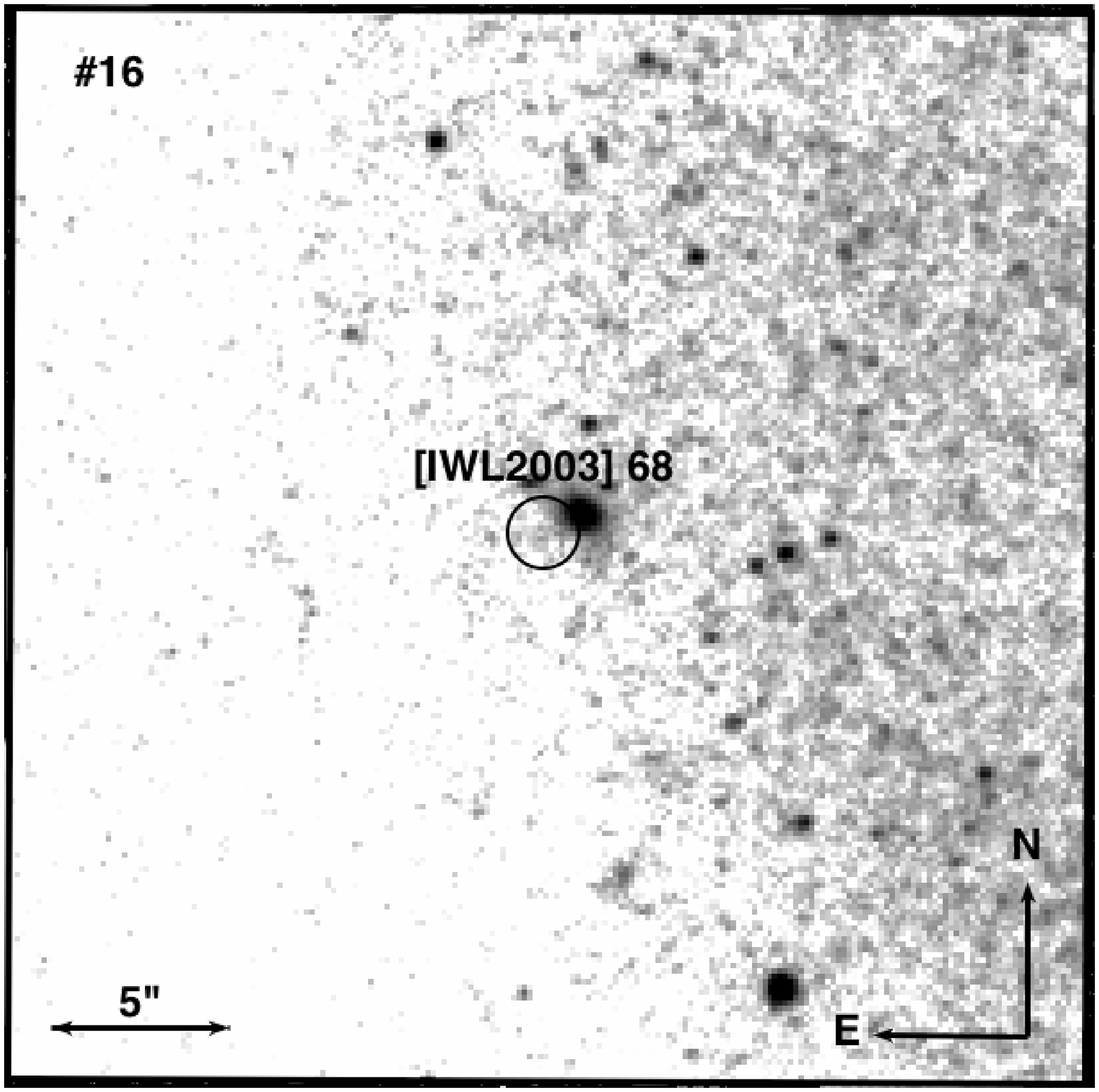}
 \includegraphics[width=0.33\textwidth]{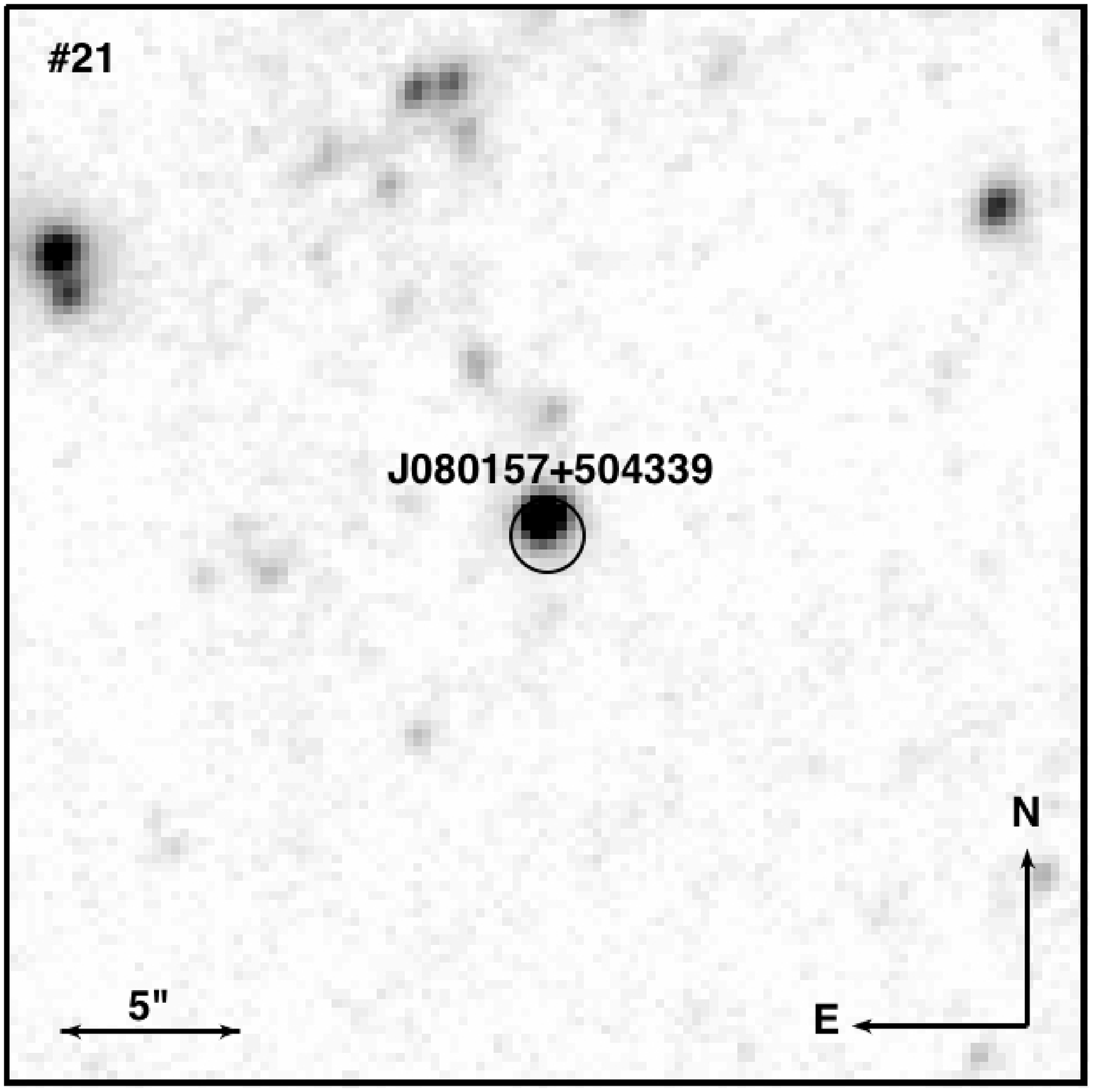}
 \includegraphics[width=0.33\textwidth]{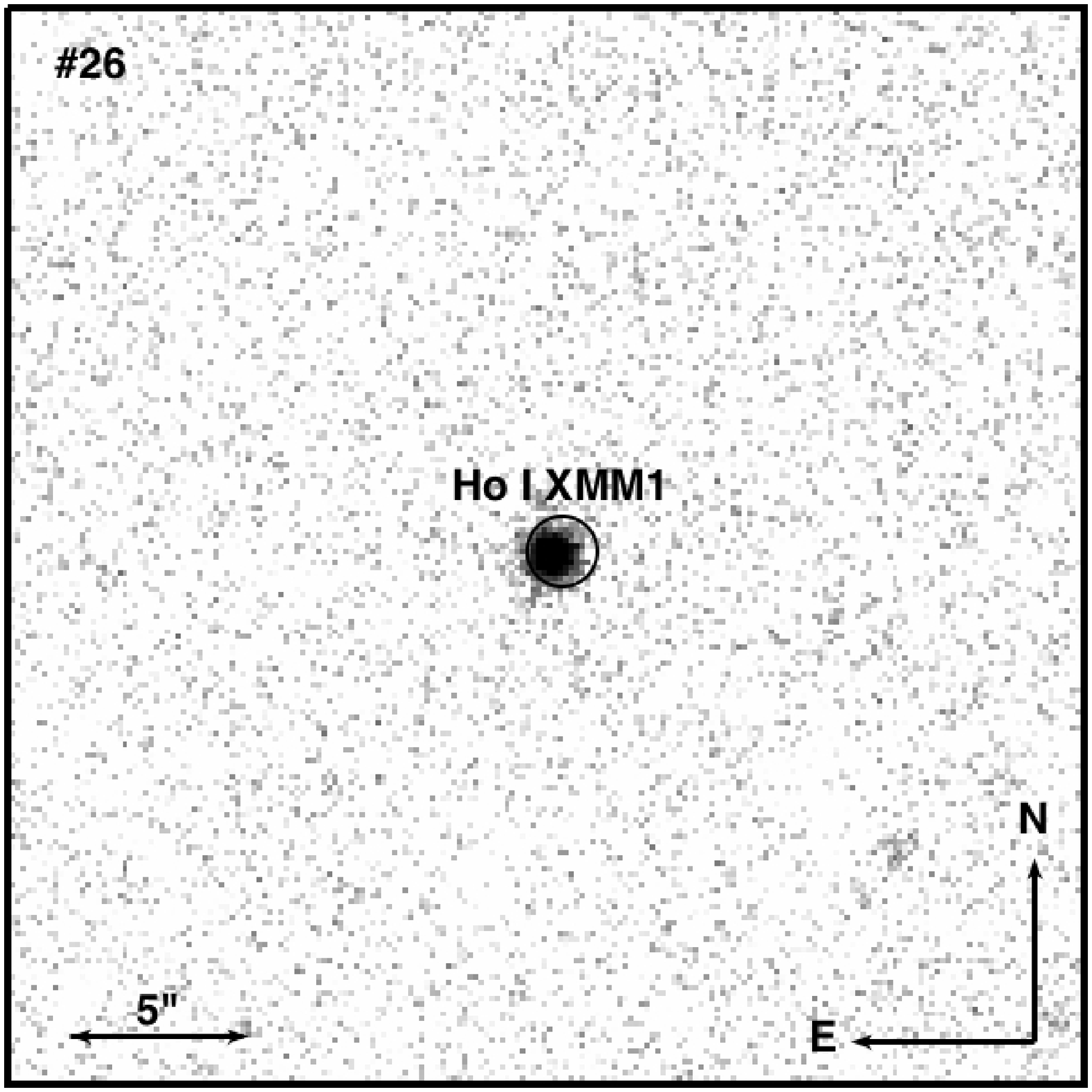}
 }
 \hbox{
\includegraphics[width=0.33\textwidth]{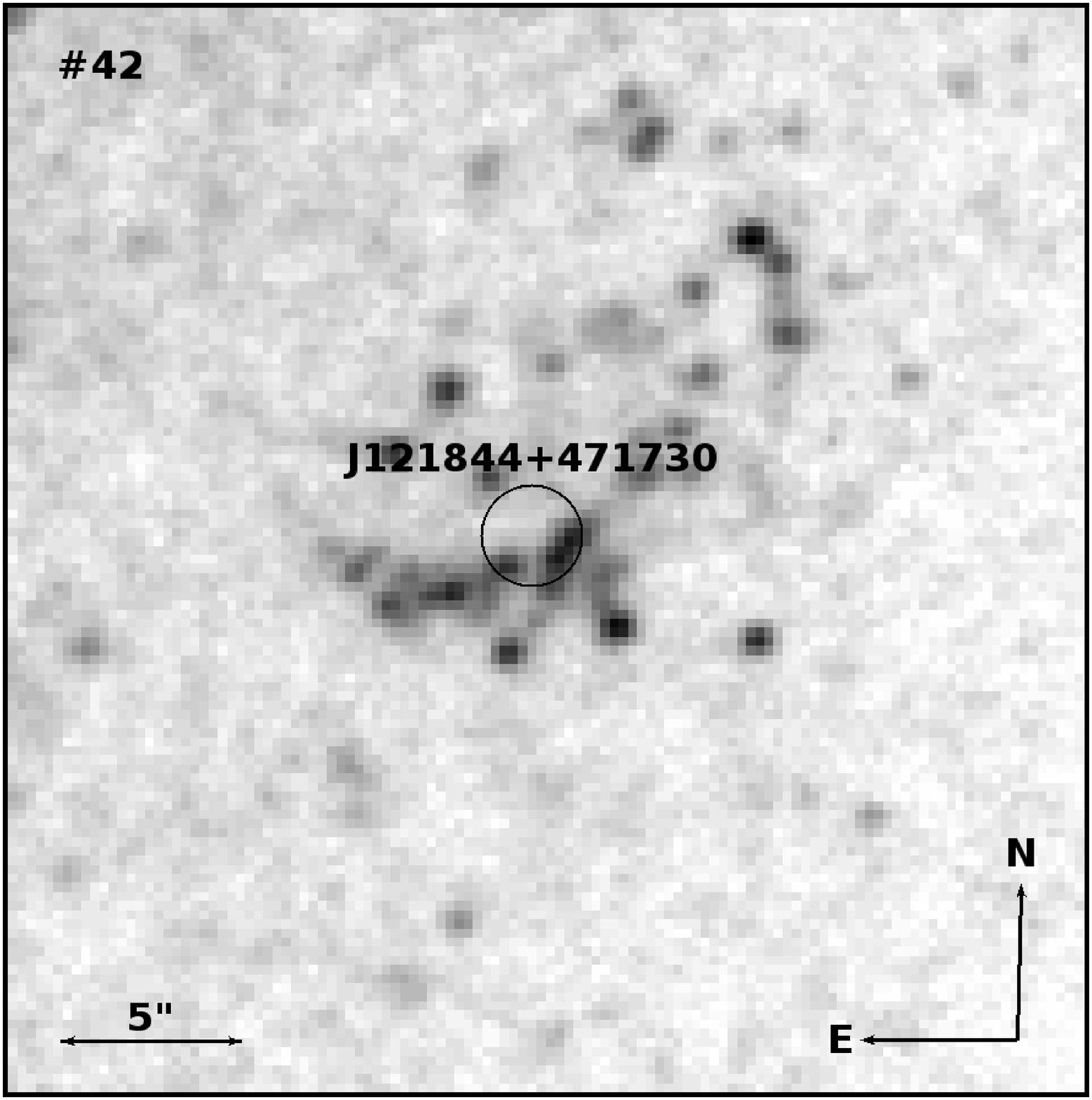}
\includegraphics[width=0.33\textwidth]{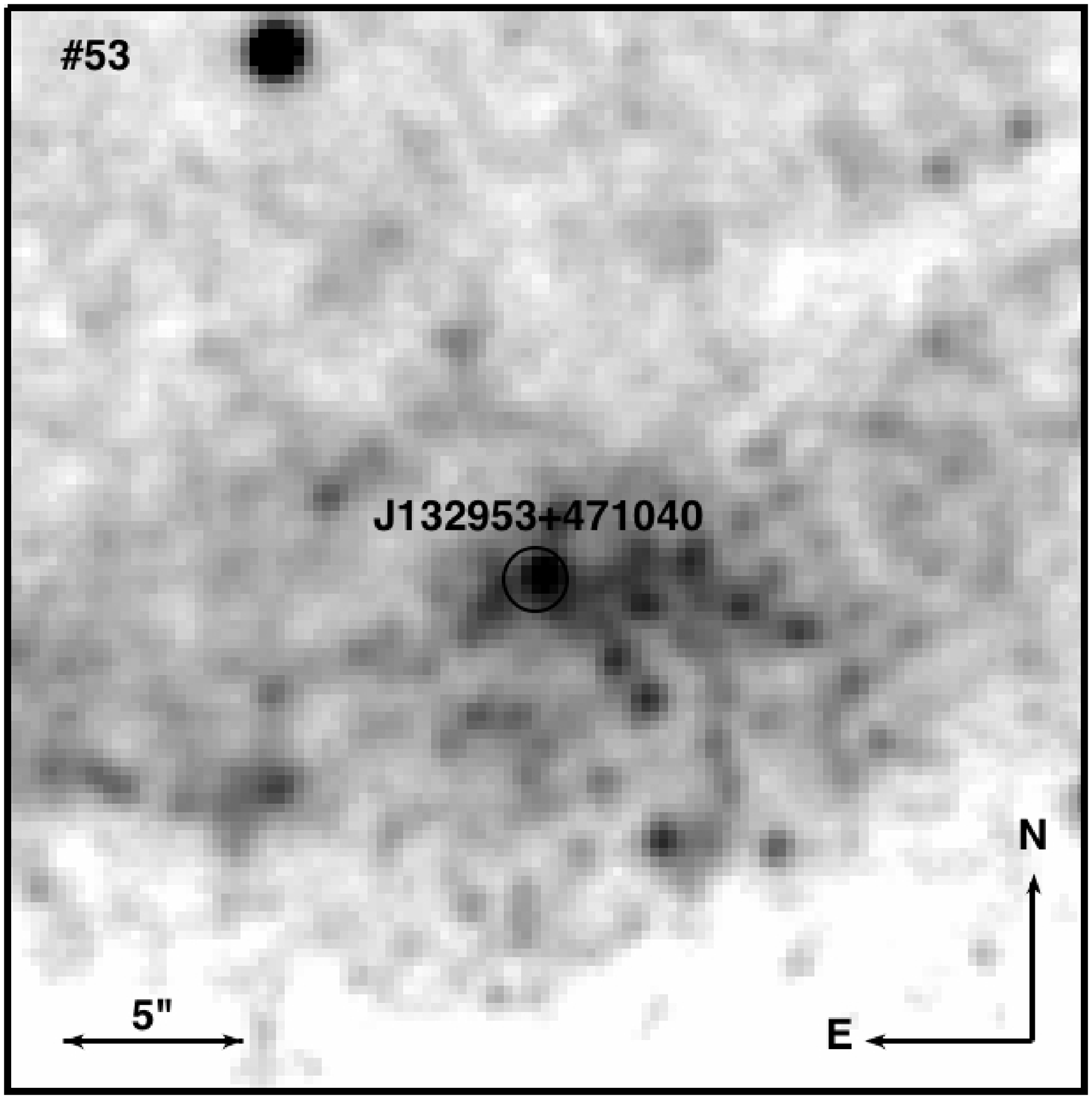}
\includegraphics[width=0.33\textwidth]{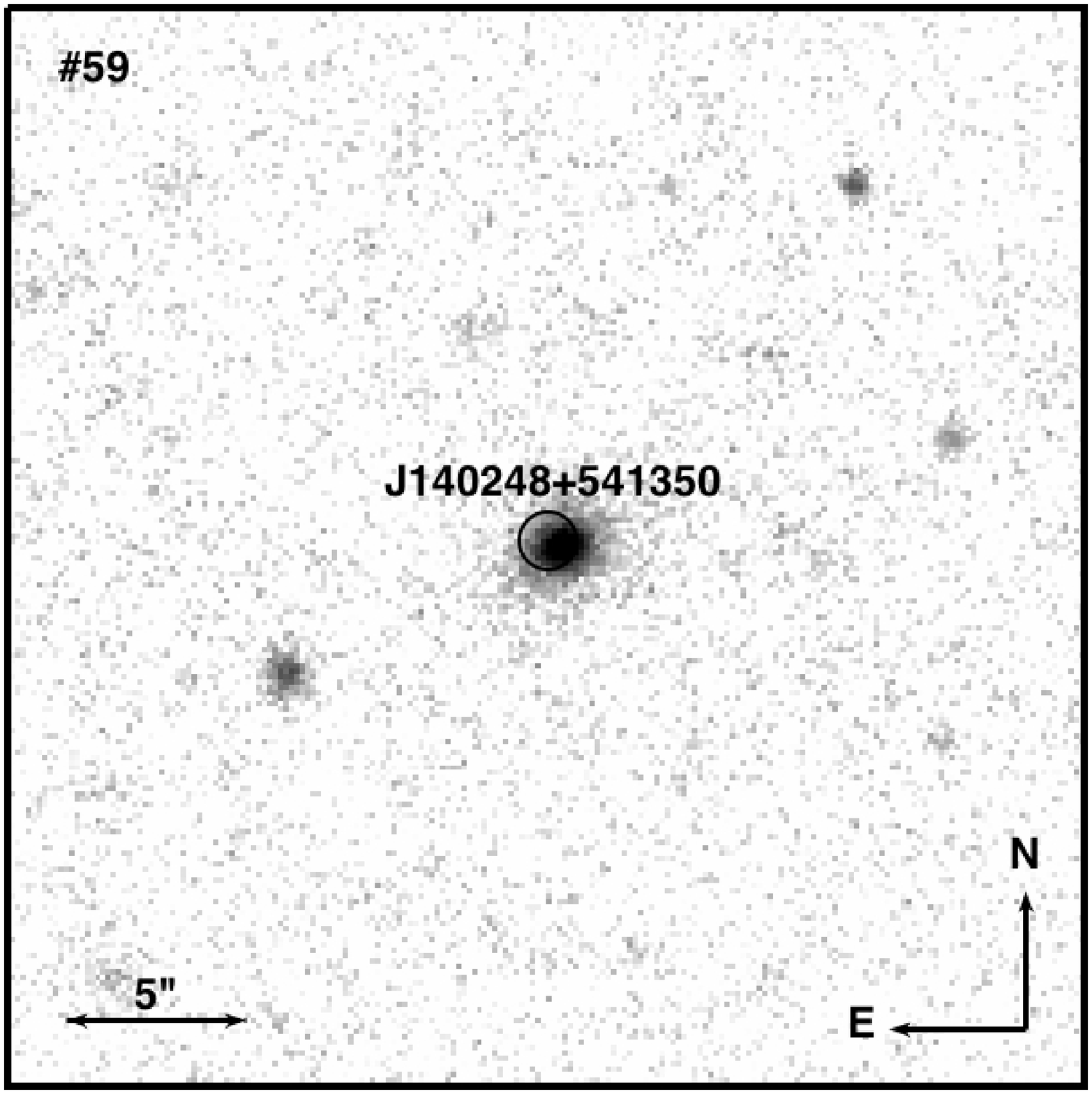}
}
 \caption{0.5' $\times$ 0.5' finder charts of all ULXs with a candidate counterpart that are likely or confirmed background AGN or (young) stellar clusters. The black circles are the 95\% confidence error circles around the positions of the ULXs. The numbers in the upper left corners refer to the image numbers in Table \ref{logtabel}.}\label{agnfinders}
\end{figure*}

\subsection{Candidate red supergiants}
For 12 sources in our sample we have detected a NIR counterpart with an absolute magnitude consistent with that of a red supergiant, one of which we suspect to be a background AGN. The NIR images of the remaining 11 RSG candidates are shown in Figure \ref{rsgfinders}. 

Two sources, J080157+504339 in NGC 2500 and Holmberg II X-1, were observed twice in the \textit{Ks}-band, J080157+504339 on January 1 and January 7, 2012 and Holmberg II X-1 on January 5, 2012 and January 27, 2013. In both cases, the source brightness was consistent with being the same at the two epochs.

We have measured $H-K$ colours for 4 of the 11 RSG candidates (J022721+333500 and J022727+333443 in NGC 925, J024323+372038 in NGC 1058 and Holmberg II X-1). The measured values are $0.7 \pm 0.3$, $0.6 \pm 0.3$, $1.1 \pm 0.5$ and $1.3 \pm 0.3$, respectively. With the caveat that the images in the two bands were taken several months apart (although if the NIR emission is indeed dominated by an RSG, it is not expected to vary significantly), and the errors are large, these are rather high values for red supergiants, whose $H-K$ colours range from 0 to 0.3 \citep{cox_book}. It is unlikely that the red colours are caused by extinction, because the hydrogen column densities needed for this would be too high. For example, to cause a reddening of $H-K = 0.5$, one needs an extinction in the $V$-band $A_V \approx 8$ (based on relative extinctions from \citealt{cardelli89}). This corresponds to a hydrogen column density $N_H \approx 1.8 \times 10^{22}$ cm$^{-2}$ \citep{guver09}. Although hydrogen column densities of ULXs can vary significantly, they do not usually reach such values. For instance for Holmberg II X-1, the hydrogen column density inferred from the X-ray spectrum is $N_H = 3.7 \times 10^{21}$ cm$^{-2}$ \citep{kaaret04}. The ULX with one of the highest absorption columns known, M82 X-1, has $N_H = 1.12 \times 10^{22}$ cm$^{-2}$ \citep{kaaret06}, still not nearly enough to account for a reddening of $H-K > 0.5$. Other possible mechanisms that could cause the red $H-K$ colours include a contribution from jets, although that seems unlikely based on the expected NIR contribution from the jet in Holmberg II X-1. It is also possible that there are strong (nebular) emission lines present that cause the red colours if those are stronger in the $K$- than in the $H$-band. 

\subsubsection*{J004722-252051}
This ULX candidate in NGC 253 has an X-ray luminosity of $(2.9 \pm 0.12) \times 10^{39}$ \lum{} (90\% confidence). The X-ray source is variable on short timescales (\citealt{barnard10}, their ULX1). We detected a counterpart to this X-ray source in our \textit{Ks}-band image with an absolute magnitude of \textit{Ks} = $-10.5 \pm 0.03 \pm 0.5 \pm 0.10$ (the errors are the statistical error on the magnitude, the error on the zeropoint of the image and the error on the distance modulus, respectively). This absolute magnitude is consistent with that of an M-type supergiant; spectroscopic observations may shed more light on the nature of this source.

\subsubsection*{J022721+333500}
We detected the counterpart to this ULX candidate in NGC 925 in both the $H$- and the \textit{Ks}- band, with absolute magnitudes $H = -10.6 \pm 0.03 \pm 0.2 \pm 0.4$ and $Ks = -11.3 \pm 0.03 \pm 0.2 \pm 0.4$. The source is rather bright for a single star, but within the uncertainties the absolute magnitudes are still consistent with an M-type supergiant. The counterpart is extended in both the $H$- and \textit{Ks}-band image (see Figure \ref{rsgfinders}). If it indeed consists of two sources of about equal brightness, one of them could be the true counterpart of the ULX and that single source would be fainter than the absolute magnitudes that we find for the combined sources.

\subsubsection*{J022727+333443}
The second ULX candidate in NGC 925 has a higher X-ray luminosity than most sources in our sample, reaching $\sim 2.5 \times 10^{40}$ \lum{} \citep{swartz11}. A candidate counterpart is detected in both our $H$- and \textit{Ks}-band image, with absolute magnitudes of \textit{Ks} = $-9.8 \pm 0.08 \pm 0.2 \pm 0.4$ and $H = -9.2 \pm 0.08 \pm 0.2 \pm 0.4$. These magnitudes are consistent with those of a red supergiant.

\subsubsection*{J024323+372038}
This ULX candidate is located in the outskirts of NGC 1058. It has a candidate counterpart with absolute magnitudes of \textit{Ks} = $-10.1 \pm 0.06 \pm 0.4 \pm 0.4$ and $H = -9.0 \pm 0.2 \pm 0.3 \pm 0.4$, if it is at the distance to NGC 1058. These absolute magnitudes are compatible with the source being a red supergiant. Because the ULX candidate is not located in a spiral arm and no signs of recent star formation are visible in the NIR images, we have to take into account the possibility that this is a foreground star or background AGN. Spectroscopic observations are necessary to determine the true distance to this source.

\subsubsection*{Holmberg II X-1}
Holmberg II X-1 is a ULX with an X-ray luminosity of $\sim 10^{40}$ \lum. It has an optical counterpart with a $V-$band magnitude of $21.86 \pm 0.09$ surrounded by an ionized nebula (\citealt{kaaret04,pakull02,moon11}) and its UV and optical emission are best fitted by an irradiated disc model \citep{tao12a}. The source has also been detected in radio observations, and recently \citet{cseh13b} reported the discovery of recurrent radio jets from the ULX. Taking the flux density of the radio core and the spectral index of $\alpha = -0.8 \pm 0.2$ reported by \citet{cseh13b}, the NIR emission expected from the central component has an apparent \textit{Ks}-band magnitude of $\sim$25.3, 6 magnitudes fainter than the counterpart that we detected. This spectrum is consistent with optically thin synchrotron emission, as expected for intermittent jets occurring at high accretion rates \citep{fender09}. 
If we assume a flat spectrum (expected for steady jets in the low/hard state) then the \textit{Ks}-band magnitude would be $\sim$16.5, 3.5 magnitudes brighter than what we detect. The spectral index needed to explain the \textit{Ks}-band emission is $\alpha \approx -0.25$. Note that the X-ray spectrum of Holmberg II X-1 indicates that it is not in the low/hard state but instead in an `ultraluminous' state, possibly accreting above the Eddington limit (e.g. \citealt{gladstone09,sutton13}). It is therefore unlikely that the radio emission is due to a steady jet with a flat spectrum and we do not expect jet emission to contribute significantly to the NIR light.

The absolute magnitude of the counterpart of \textit{Ks} = $-8.35 \pm 0.08 \pm 0.10 \pm 0.03$, $H = -7.1 \pm 0.3 \pm 0.10 \pm 0.03$ indicates that the infrared excess might be due to a red supergiant companion. It could also be related to the nebula surrounding the ULX.

\subsubsection*{J112018+125900}
J112018+125900 is a ULX candidate located on the edge of a spiral arm in NGC 3627. Its absolute \textit{Ks}-band magnitude of $-9.1 \pm 1.9 \pm 0.7 \pm 0.4$ puts it in the magnitude range of red supergiants, but its apparent magnitude ($20.6 \pm 1.9 \pm 0.7$ in the \textit{Ks}-band) and its location in a crowded region will make it difficult to observe it spectroscopically with current instrumentation.

\subsubsection*{J120922+295551}
This ULX candidate in NGC 4136 was detected by the ROSAT High Resolution Imager (HRI) at $2.5 \times 10^{39}$ \lum{} \citep{lira00}, and subsequently at much lower luminosity in a \chan{} observation by \citet{roberts04}. We detected a counterpart with an absolute $H$-band magnitude of $-10.78 \pm 0.03 \pm 0.10 \pm 0.4$. Within the uncertainties this is compatible with an M-type RSG companion.

\subsubsection*{J120922+295559}
The second ULX candidate in NGC 4136 was discovered by \citet{roberts04} in a \chan{} observation, at $\sim 11''$ from J120922+295551 with a maximum unabsorbed X-ray luminosity of $2.6 \times 10^{39}$ \lum. Its NIR counterpart has an absolute $H$-band magnitude of $-10.75 \pm 0.03 \pm 0.10 \pm 0.4$. Within the uncertainties this is compatible with an M-type RSG companion.

\subsubsection*{NGC 5408 X-1}
\citet{lang07} discovered an optical counterpart to this ULX in \textit{Hubble Space Telescope} (\textit{HST}) images with a $V$- and $I$-band magnitude of 22.4. The counterpart has also been studied by \citet{grise12}, who discovered it in 6 \textit{HST} filters from the near UV to the NIR. They find that the flux of the source drops continuously with increasing wavelength to $\sim 0.13 \times 10^{-18}$ erg s$^{-1}$ cm$^{-2}$ \AA{}$^{-1}$ in the $H$-band and concluded that the companion star was either a blue (O- or B-type) supergiant or the optical emission was dominated by the accretion disc. Optical spectra obtained with the VLT show a blue continuum superimposed with emission lines from the surrounding nebula (\citealt{kaaret09,cseh11}). 

Radio emission from the nebula has also been detected by \citet{lang07}; since the emission is resolved in their VLA images, they conclude that it can not originate in a relativistic jet. The flux density and spectral index of this radio source are similar to those of the compact radio jet in Holmberg II X-1, so assuming that the spectrum continues into the NIR we would expect a similar magnitude for NGC 5408 X-1 of \textit{Ks} $\sim 25$, five magnitudes fainter than what we detect.

The error circle in our image actually contains two sources, but comparison with the \textit{HST} image of \citet{lang07} shows that the most Eastern source (indicated with an arrow in Figure \ref{rsgfinders}) is the counterpart. This source has an apparent \textit{Ks}-band magnitude of $20.3 \pm 0.13 \pm 0.2$. At the distance of NGC 5408 this corresponds to an absolute \textit{Ks}-band magnitude of $-8.1 \pm 0.13 \pm 0.2 \pm 0.8$. Interestingly, this corresponds to a higher flux ($\sim 0.3 \times 10^{-18}$ erg s$^{-1}$ cm$^{-2}$ \AA{}$^{-1}$) than \citet{grise12} measured in the $H$-band, while none of their models predict an increase in flux towards the \textit{Ks}-band. It is possible that this excess infrared radiation is emitted by the companion star, while most of the optical emission originates from the disc. However, RSGs typically have $H - K \approx 0$, while for this source we find $H - K \approx 2$ (based on the $H$-band magnitude reported by \citealt{grise12}). Since the \textit{Ks}-band image 
was not taken simultaneously with the optical and $H$-band data, 
variability of the source is also a possible explanation.

\subsubsection*{J1402+5440}
One of the four ULX candidates that we observed in M101, this source has a reported X-ray luminosity of $2.4 \times 10^{39}$ \lum{} \citep{swartz11}. The counterpart that we detected in our $H$-band image has an absolute magnitude of $-9.7 \pm 0.04 \pm 0.2 \pm 0.05$. This places it in the magnitude range of red supergiants.

\subsubsection*{J140314+541807}
J140314+541807 is another ULX in M101 with an X-ray luminosity of $2.9 \times 10^{39}$ \lum{} as reported by \citet{winter06}. We detected a counterpart in the $H$-band with an absolute magnitude of $-10.69 \pm 0.03 \pm 0.10 \pm 0.05$. Within the uncertainties this is compatible with an M-type RSG companion.

\subsection{Candidate star clusters and background AGN}
Five of the candidate counterparts that we discovered have absolute magnitudes that are incompatible with them being single stars. Some of these can be classified as background AGN; others might be bona fide ULXs located in stellar clusters (cf \citealt{voss11,jonker12}). Additionally, two of the ULX candidates near Holmberg I are likely not real ULXs but background AGN. We discuss these sources in more detail below; the NIR images of these sources are shown in Figure \ref{agnfinders}.

\subsubsection*{[IWL2003] 68}
The absolute magnitude of the \textit{Ks}-band counterpart to this ULX candidate in NGC 1637 is $-13.7 \pm 0.005 \pm 0.5 \pm 0.4$. This is much brighter than a single red supergiant. The source is resolved in our \textit{Ks}-band image. The counterpart has also been detected in \textit{HST} images of the galaxy by \citet{immler03b}, who report apparent magnitudes $m_V \sim 22.8$ and $m_I \sim 21.1$ and a full width at half maximum (FWHM) of 0.45'', more than twice as large as the size of the point sources in the image. At the distance to NGC 1637 this corresponds to a physical size of $\sim 25$ pc. The counterpart could be a star cluster in NGC 1637, or a background galaxy containing an AGN. 

\subsubsection*{J080157+504339}
This ULX candidate, situated in NGC 2500, has a bright counterpart that we detected in two \textit{Ks}-band observations at an absolute magnitude of $-14.3 \pm 0.002 \pm 0.2 \pm 0.4$ and $-14.1 \pm 0.005 \pm 0.15 \pm 0.4$, respectively, if the distance to NGC 2500 is assumed. However, optical spectroscopic observations by \citet{gutierrez13} have shown this source to be a background AGN.

\subsubsection*{Ho I XMM1}
The ULX candidate Ho I XMM1 is situated at 5' from the center of the dwarf irregular galaxy Holmberg I. The galaxy has a diameter (semimajor axis) of 3.6' \citep{gildepaz07}; the X-ray source is thus located outside the galaxy itself. The $H$-band counterpart that we detected has an absolute magnitude of $-10.14 \pm 0.01 \pm 0.10 \pm 0.03$. This in itself does not exclude the possibility that the counterpart is a red supergiant. However, it has also been detected by the Wide-Field Infrared Survey Explorer (WISE) and its colours in the WISE bands are [3.4]-[4.6] = 0.8, [4.6]-[12] = 2.5, placing it firmly among the AGN in WISE colour-colour diagrams (cf \citealt{stern05,dabrusco12}). Thus it is likely that this ULX candidate is a background AGN like J080157+504339 in NGC 2500.

\subsubsection*{J0940+7106}
J0940+7106 is another ULX candidate 5' from the center of Holmberg I. It does not have a counterpart within the 95\% confidence error circle of 1.1'', but a single NIR point source is located just outside it. This NIR source was also detected by WISE. Its colours are [3.4]-[4.6] = 1.35, [4.6]-[12] = 2.4, which place the source among the Seyfert galaxies in the diagram of \citet{dabrusco12}. Therefore we expect that the X-ray source and the NIR source are related, and that this ULX candidate is also in fact a background AGN.

\subsubsection*{J121844+471730}
This ULX candidate in NGC 4258 is located in the center of a cluster of stars. The counterpart has an absolute $H$-band magnitude of $-11.50 \pm 0.02 \pm 0.1 \pm 0.02$ and is thus too bright to be a single star. 

\subsubsection*{J132953+471040}
The counterpart to J132953+471040 in M51 has an absolute $H$-band magnitude of $-13.88 \pm 0.02 \pm 0.10 \pm 0.2$. Comparing to the \textit{HST} image of \citet{terashima06}, we find that this source is at the center of the star cluster in which the ULX is located. Even in the higher resolution \textit{HST} image this source is unresolved, although extended, and it is not possible to distinguish single stars. 

\subsubsection*{J140248+541350}
This ULX candidate in M101 is associated with an optical counterpart in the Sloan Digital Sky Survey (SDSS) catalog \citep{pineau11}. The SDSS source is classified as a galaxy in all Sloan filters except the $u'$-band. We also detected the counterpart in our $H$-band image, where it is clearly extended (see Figure \ref{agnfinders}). This source is likely to be a background galaxy harbouring an AGN.

\subsection{Non-detections of sources with known optical counterparts}
Several ULX candidates in our sample have known optical counterparts, and optical spectra are available for a few. Some of these we did detect in our NIR images (see Sections 5.1 and 5.2), but the majority were not detected by us. \citet{ptak06}, \citet{tao11} and \citet{gladstone13} found (sometimes multiple) candidate optical counterparts in archival \textit{HST} data for 11 ULXs that we do not detect in our images (J004742-251501, J0318-6629, J034555+680455, J034615+681112, J073625+653539, NGC 3031 ULX1, Holmberg IX X-1, J112020+125846, IXO 53, J123551+27561, J132519-430312 and J132938+582506). Optical spectra are available for Holmberg IX X-1, J132938+582506 in NGC 5204 (both in \citealt{roberts11}) and J140332+542103 in M101 \citep{liu13}. 

The $V$-band apparent magnitudes for these counterparts lie between 22 and 25. Holmberg II X-1 and NGC 5408 X-1, the only two RSG candidates that we detect in the NIR that have also been detected in the optical, have values for $V - K \sim 2.1-2.5$ (not taking into account possible variability). Typical $V - K$ values for RSGs range from $\sim 2$ to 4, but in ULXs there will also be a contribution from the accretion disc to the optical light, which lowers $V - K$. 
For the ULXs with $V-$band magnitude up to $\sim 22$ we can state that they have $V - K$ closer to 0, because otherwise we would have detected them in our NIR images. Indeed, for Holmberg IX X-1 (with a $V-$band magnitude of $\sim 22.8$, \citealt{grise06}) there is also an \textit{HST} $H$-band observation in which the counterpart is detected at a magnitude of 22.56 (Gris\'e et al., in prep.). These $V - K$ values are in line with what would be expected for early type companion stars \citep{elias85}. The ULXs with $V \approx 23-24$ we would not have detected in our NIR images if they have $V - K \approx 2.5$, but we can exclude that they have $V - K$ values that are even higher. The ULXs with the very faintest optical counterparts, with $V \gtrsim 25$, we could not detect even if they had $V - K \approx 4$.

\section{Conclusions}
We have performed the first systematic search for near-infrared counterparts to nearby ULXs. We observed 62 ULXs in the $H$- and/or \textit{Ks}-band and detected candidate counterparts for 17 of them. For the other 45 ULXs we determined limiting magnitudes. Of the 17 ULXs with NIR counterparts, 11 had no previously reported optical or NIR counterpart. We detected 11 candidate counterparts with absolute magnitudes that are consistent with them being single red supergiants. Two of these (Holmberg II X-1 and NGC 5408 X-1) also have known optical counterparts. Holmberg II X-1 has a radio jet, but the NIR emission that we detect is 6 magnitudes brighter than the emission expected from the jet. In NGC 5408 X-1, we detected excess \textit{Ks}-band emission compared to the models tested by \citet{grise12}. This excess infrared radiation could be emitted by the companion star, while most of the optical emission originates from the disc. However this would not explain the large value for $H - K$ in this ULX. It is also possible that the source is variable.

Six counterparts are too bright to be single stars and in some cases extended; they are likely star clusters or background galaxies. 

The fact that we detect only a fraction of the ULXs in our NIR images points towards differences between ULX systems: the systems that show relatively strong NIR emission might have larger accretion discs than the ones that do not, or there may be strong nebular lines present that are absent in other sources. Alternatively, these systems may contain red supergiant donor stars that are intrinsically bright in the NIR. If the NIR emission indeed originates from a late-type donor star, these systems are excellent candidates for future spectroscopic studies to find dynamical masses for their black holes.

\section*{Acknowledgements}
EK thanks Mischa Schirmer for his help with the data reduction software {\sc Theli}. We thank the anonymous referee for their comments that helped improve the paper. TPR's contribution to this paper was funded as part of STFC consolidated grant ST/K000861/1. Based on observations collected at the European Southern Observatory, Chile, programmes 089.D-0663(A) and 090.D-0417(A). The William Hershel Telescope is operated on the island of La Palma by the Isaac Newton Group in the Spanish Observatorio del Roque de los Muchachos of the Instituto de Astrofi­sica de Canarias. Observations reported here were obtained at the MMT Observatory, a joint facility of the Smithsonian Institution and the University of Arizona. 
\bibliographystyle{mn_new}
\bibliography{bibliography}

\end{document}